\documentclass[twocolumn]{aastex631}

\usepackage{mathrsfs}
\usepackage{threeparttable}
\begin{document}
\title{Quasi-periodic X-ray eruptions and tidal disruption events prefer similar host galaxies}
\correspondingauthor{Thomas Wevers}
\email{twevers@stsci.edu}
\author[0000-0002-0786-7307]{T. Wevers}
\affiliation{Space Telescope Science Institute, 3700 San Martin Drive, Baltimore, MD 21218, USA}
\author[0000-0002-4235-7337]{K. D. French}
\affiliation{Department of Astronomy, University of Illinois, 1002 W. Green St., Urbana, IL 61801, USA}
\author{A. I. Zabludoff}
\affiliation{Department of Astronomy and Steward Observatory, University of Arizona, 933 N Cherry Ave, Tucson, AZ 85721, USA}
\author{T. Fischer}
\affiliation{AURA for ESA, Space Telescope Science Institute, 3700 San Martin Drive, Baltimore, MD 21218, USA}
\author{K. Rowlands}
\affiliation{AURA for ESA, Space Telescope Science Institute, 3700 San Martin Drive, Baltimore, MD 21218, USA}
\affiliation{William H. Miller III Department of Physics and Astronomy, Johns Hopkins University, Baltimore, MD 21218, USA}
\author{M. Guolo}
\affiliation{Bloomberg Center for Physics and Astronomy, Johns Hopkins University, 3400 N. Charles St., Baltimore, MD 21218, USA}
\author{B. Dalla Barba}
\affiliation{Università degli studi dell’Insubria, Via Valleggio 11, Como 22100, Italy}
\affiliation{Osservatorio Astronomico di Brera, Istituto Nazionale di Astrofisica (INAF), Via E. Bianchi 46, Merate (LC) 23807, Italy}
\author{R. Arcodia}
\affiliation{MIT Kavli Institute for Astrophysics and Space Research, 70 Vassar Street, Cambridge, MA 02139, USA}
\author{M. Berton}
\affiliation{European Southern Observatory, Alonso de Córdova 3107, Vitacura, Santiago, Chile}
\author{F. Bian}
\affiliation{European Southern Observatory, Alonso de Córdova 3107, Vitacura, Santiago, Chile}
\author[0000-0002-8304-1988]{I. Linial}
\affiliation{Institute for Advanced Study, 1 Einstein Drive, Princeton, NJ 08540, USA}
\affiliation{Department of Physics and Columbia Astrophysics Laboratory, Columbia University, New York, NY 10027, USA}
\author{G. Miniutti}
\affiliation{Centro de Astrobiología (CAB), CSIC-INTA, Camino Bajo del Castillo s/n, ESAC campus, 28692 Villanueva de la Cañada, Madrid, Spain}
\author[0000-0003-1386-7861]{D.R. Pasham}
\affiliation{MIT Kavli Institute for Astrophysics and Space Research, 70 Vassar Street, Cambridge, MA 02139, USA}

\begin{abstract}
In the past five years, six quasi-periodic X-ray eruption (QPE) sources have been discovered in the nuclei of nearby galaxies. Their origin remains an open question.
We present MUSE integral field spectroscopy of five QPE host galaxies to characterize their properties.
We find that 3/5 galaxies host extended emission line regions (EELRs) up to 10 kpc in size. The EELRs are photo-ionized by a non-stellar continuum, but the current nuclear luminosity is insufficient to power the observed emission lines. The EELRs are decoupled from the stars both kinematically and in projected sky position, and the low velocities and velocity dispersions ($<$ 100 km s$^{-1}$ and $\lesssim 75$ km s$^{-1}$ respectively) are inconsistent with being AGN- or shock-driven. The origin of the EELRs is likely a previous phase of nuclear activity.
The QPE host galaxy properties are strikingly similar to those of tidal disruption events (Wevers et al. submitted). The preference for a very short-lived (the typical EELR lifetime is $\sim$15\,000 years), gas-rich phase where the nucleus has recently faded significantly suggests that TDEs and QPEs may share a common formation channel, disfavoring AGN accretion disk instabilities as the origin of QPEs. In the assumption that QPEs are related to extreme mass ratio inspiral systems (EMRIs; stellar-mass objects on bound orbits about massive black holes), the high incidence of EELRs and recently faded nuclear activity can be used to aid in the localization of the host galaxies of EMRIs discovered by low frequency gravitational wave observatories.
\end{abstract}

\section{Introduction} \label{sec:intro}
Quasi-periodic X-ray eruptions (QPEs) are a recent addition to the various modes of rapid variability observed in massive black holes (MBHs) inhabiting galactic nuclei. Their timing and spectral properties, including quasi-periodic behavior and the emergence of an additional hot thermal component during the outburst rise, are unique among the known variability of active galactic nuclei (AGN; \citealt{Miniutti19, Giustini20, Arcodia21}). QPEs may relate to accretion disk instabilities \citep{Sniegowska20, Raj21, Pan23} or to the interaction between the SMBH (or an accretion disk surrounding it) and a stellar-mass companion. The latter class of models comes in many flavors, including repeated partial tidal disruptions \citep{King22}, Roche-lobe overflow \citep{Lu22, Krolik22, Metzger22} and star- or BH-disk interactions \citep{Linial23, Franchini23, Tagawa23,Zhou24}. 

Observationally, several QPEs occur over the course of a long-term ($\sim$years) lightcurve decline, which may be related to the tidal disruptions of stars \citep{Miniutti23, Arcodia24}. Some QPEs also exhibit rebrightening episodes. Other clues that hint at a potential connection between QPEs and tidal disruption events (TDEs) include the discovery of QPE-like flares in TDE candidates \citep{Chakraborty21, Quintin23}. From a theoretical perspective, some models require a compact accretion disk to explain the QPEs, which similarly require a TDE to render the QPEs visible \citep{Linial23}. 

Further similarities between QPEs and TDEs can be found in their host galaxies, which have been inferred to contain black holes smaller ($\sim$10$^6$ M$_{\odot}$) than in typical AGNs. \citet{Wevers22} also noted that QPEs are overrepresented among post-starburst (PSB) and quiescent Balmer strong (QBS) galaxies, which are rare ($\lesssim$0.2 and 2 per cent, respectively) in the local Universe. TDEs are similarly known to be overrepresented among PSB and QBS galaxies \citep{French16, Graur18}. A recent study of integral field spectroscopy of a large sample of PSB galaxies by \citet{French23} revealed that of the six sources that have extended emission line regions (EELRs) of ionized gas, one was the host galaxy of a TDE (AT2019azh) and one was host to a QPE (RXJ1301). Evidence for an EELR also exists in {\it Hubble Space Telescope} (HST) narrow-band imaging of the host galaxy of the QPE GSN069, including a marginally resolved, compact ($<$ 35 pc) component and a larger scale ($\sim$2 kpc, all throughout the HST field of view) component \citep{Patra24}. 

Based on the presence of line ratios that require a hard, non-stellar continuum in long-slit spectroscopy of the QPE nuclei, \citet{Wevers22} suggested that a long-lived accretion flow may play an important role in the QPE phenomenon. A study of the spatially resolved properties of the QPE host galaxies can be used to trace their merger and accretion history, which can provide further clues to the importance (or lack thereof) of AGN activity on the rate of nuclear transients such as QPEs and TDEs. 

In this work we present MUSE integral field spectroscopy of five QPE host galaxies. We describe the observations and data analysis in Section \ref{sec:analysis}. We present the main results in Section \ref{sec:results}, including the presence of extended emission line regions of ionized gas with peculiar kinematics that are decoupled from the stellar motions, and evidence for recently faded AGNs. We discuss these results in the context of the connection between QPEs and TDEs based on their similar host galaxy preferences, as well as implications for the theoretical models invoked to explain QPEs, in Section \ref{sec:discussion}, and present the main conclusions in 
Section \ref{sec:summary}.

\section{Observations and Analysis}
\label{sec:analysis}
We used the Multi Unit Spectroscopic Explorer (MUSE, \citealt{Bacon_2010}) instrument mounted on the Very Large Telescope Unit 4 (Yepun) to observe the host galaxies of 5/6 known QPE sources \citep{Miniutti19, Giustini20, Arcodia21, Arcodia24}. MUSE integral field observations cover a 1$\times$1 arcmin field of view with a spatial sampling (spaxel size) of 0.2 arcsec. The approximately constant spectral resolution of $\approx$2.62\AA\ corresponds to a velocity resolution of 80 (160) km s$^{-1}$ in the blue (red) wavelength range. An overview of the basic source properties can be found in Table \ref{tab:observations}, and a full observing log is provided in Table \ref{tab:muse}.

Galactic foreground extinction is removed prior to the data analysis in accordance with the E(B--V) values reported by \citet{Schlafly2011} as tabulated in Table \ref{tab:observations} and assuming R$_V$ = 3.1. 

We use pseudo-aperture spectra (with a size equal to the typical PSF FWHM of the observations in Table \ref{tab:muse}) of the nuclei to study the nuclear stellar properties. We also analyze the individual spaxel data where the signal to noise ratio around 5400 \AA\ is $>$5, and we create pseudo-aperture spectra in various regions of the extended emission line regions to confirm the results of our spatially resolved analysis with higher SNR. 

To analyze the morphology, stellar and ionized gas kinematics and properties we use the penalized pixel fitting routine (PPXF; \citealt{Cappellari23}). All quoted line velocity dispersions are corrected for instrumental broadening.

\subsection{Kinematics and line fluxes}
We employ the full MILES single stellar population (SSP) library \citep{Vazdekis2015} to fit the stellar continuum contribution as a linear combination of SSPs and constrain the stellar kinematics. Regions with strong emission and (telluric) absorption lines (as well as the Na\,\textsc{i} D doublet) are masked during this process. 

Following the determination of the best-fit SSP templates, they are subtracted before analyzing the emission lines. There is no evidence of broad emission lines in the galaxy nuclei (see also \citealt{Wevers22}), nor for multiple kinematic components to the narrow lines (see Figures \ref{fig:spectra_gsn069}, \ref{fig:spectra_rxj1301} and \ref{fig:spectra_qpe2} for example spectra). We therefore use a single kinematic component, parameterized by a narrow Gaussian, to determine the velocity dispersions, velocities and fluxes of the emission lines on the scale of the native spaxel size. Note that the PSF FWHM is generally much larger (see Table \ref{tab:muse}), so at the smallest spatial scales these measurements will be correlated. 
We have verified that decoupling the line widths and velocities for each emission line (or doublet) does not significantly alter the results. Similarly, the stellar continuum subtraction does not significantly alter the inferred gas kinematics because the EELRs are generally isolated from the continuum emission.

Three-color composite images including the emission line maps and a continuum band image are shown in Figure \ref{fig:rgb} to visualize the resulting flux distributions and to study the ionized gas morphology. 

\subsection{Star formation histories}
Two of the host galaxies (RXJ1301 and eRO-QPE1) are spectroscopically classified as PSB (Lick index H${\delta_{A}} > 4$ \AA\ and EW(H$\alpha$) $<$3 \AA) or quiescent Balmer strong (QBS, H${\delta_{A}} > 1.31$ \AA) galaxies \citep{Wevers22}. In addition, the MUSE data reveal that GSN069 and eRO-QPE2 show clear signs of ongoing star formation: GSN069 has a bright circumnuclear star-forming ring, and both GSN069 and eRO-QPE2 show several star forming knots tracing spiral arms (Figure \ref{fig:rgb}). eRO-QPE3 is in the star-forming region of the BPT and WHAN diagrams, although the current star formation rate is likely low (Fig. \ref{fig:qpe3}).

We constrain the star formation history (SFH) by assuming that each galaxy experienced a recent burst of star formation. We constrain the post-starburst age (age since 90$\%$ of the burst stars formed) and mass of new stars that formed by modeling the integrated nuclear spectra with Bagpipes \citep{Carnall2018, Carnall2019}. We assume a SFH consisting of an old delayed exponential and a young burst exponential, using a similar method as \citet{French23}. We allow the masses and ages of both components to vary, as well as the duration of the young burst component, with logarithmic priors. We assume a \citet{Kroupa2001} initial mass function and a Calzetti dust law, allowing the attenuation to vary. Following \citet{Carnall2019}, we include three terms in a Gaussian process noise model. 

We find that when allowing for both a young and an old burst, all galaxies prefer the inclusion of a recent burst of star formation within the last 0.2-1.3 Gyr. For RXJ1301 and eRO-QPE2, our modeling prefers a sizeable recent burst (Table \ref{tab:eelr}). This burst has been quenched for RXJ1301 (hence its PSB classification), while the burst is still ongoing for eRO-QPE2 (hence its SF/AGN classification, see \S \ref{sec:results}). eRO-QPE1 had a weaker burst about a Gyr ago and has remained quenched, leading to its QBS classification\footnote{We assume a specific SFH with a young and an old burst, but note that we do not explore other SFH scenarios which may also lead to a QBS classification without requiring a starburst.}. The burst in eRO-QPE3 is smaller still ($\sim 1-2 \%$ of stars formed). Finally, the SFH of GSN069 resembles that of eRO-QPE2, including a smaller burst fraction (5\% of stars formed) where the SF has not yet been fully quenched.

\section{Results}
\label{sec:results}
\begin{figure*}
    \centering
    \includegraphics[width=0.32\textwidth]{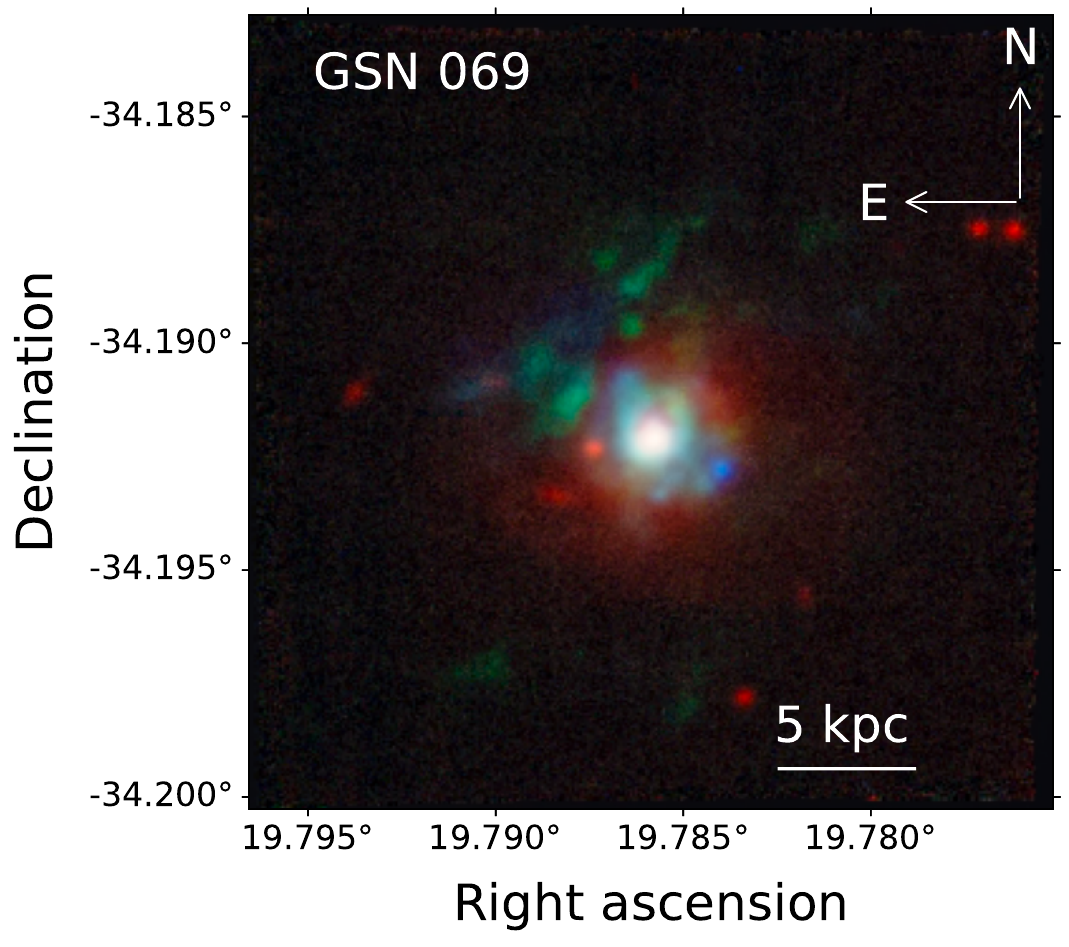}\,
    \includegraphics[width=0.32\textwidth]{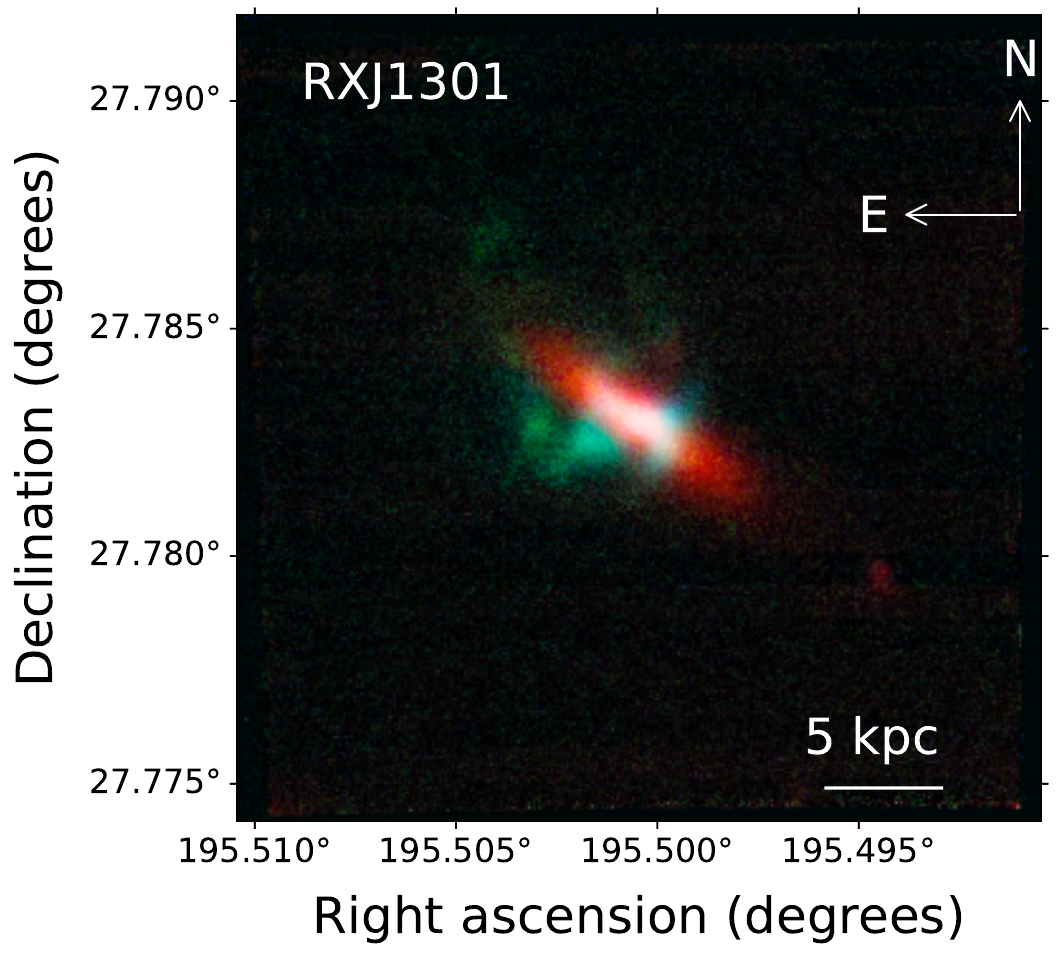}\includegraphics[width=0.32\textwidth]{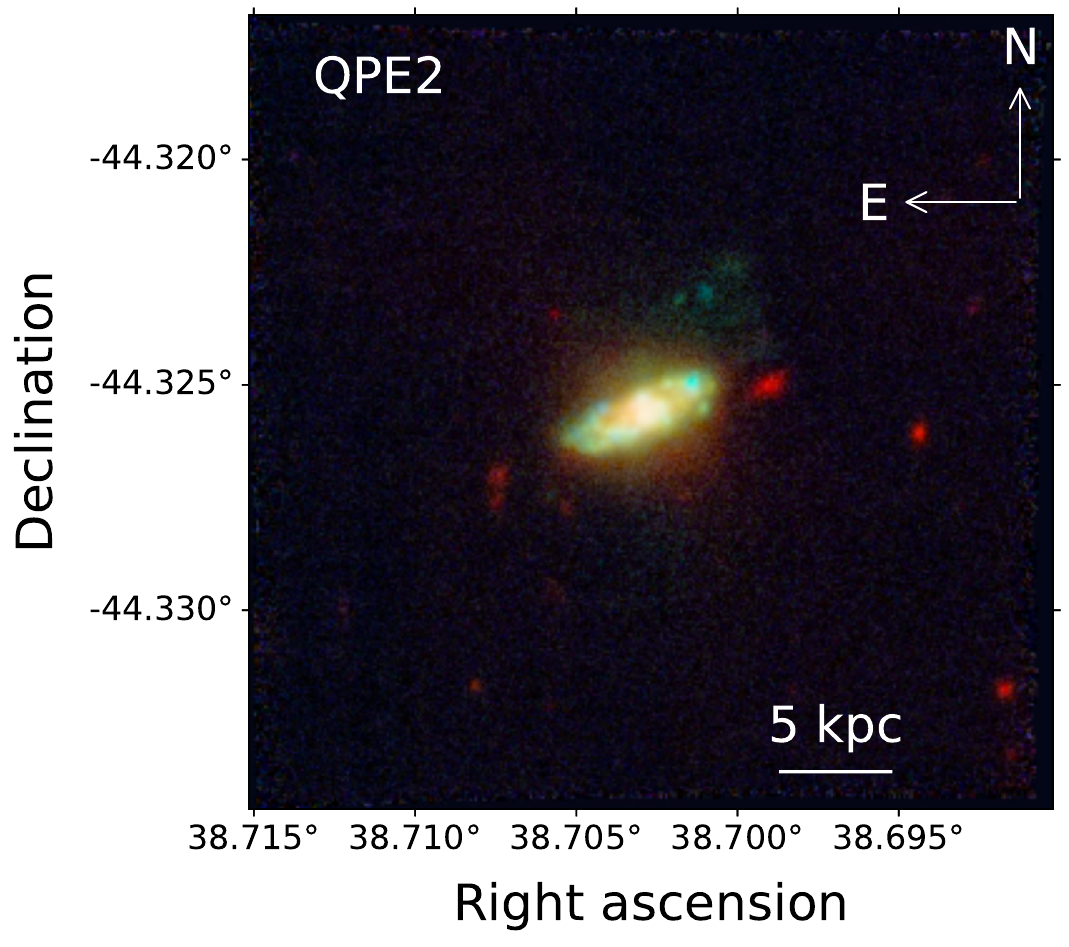}
    \includegraphics[width=0.32\textwidth]{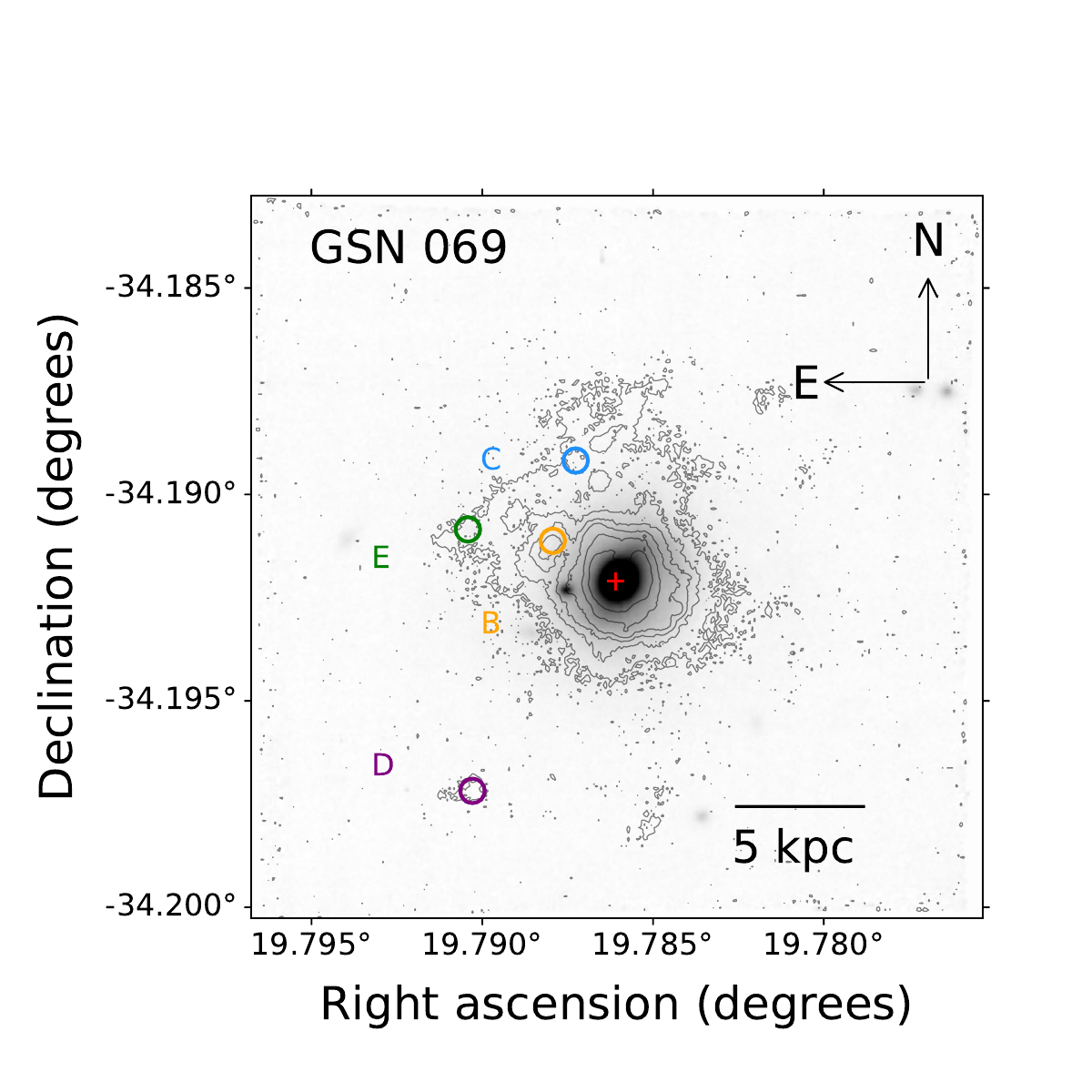}
    \includegraphics[width=0.32\textwidth]{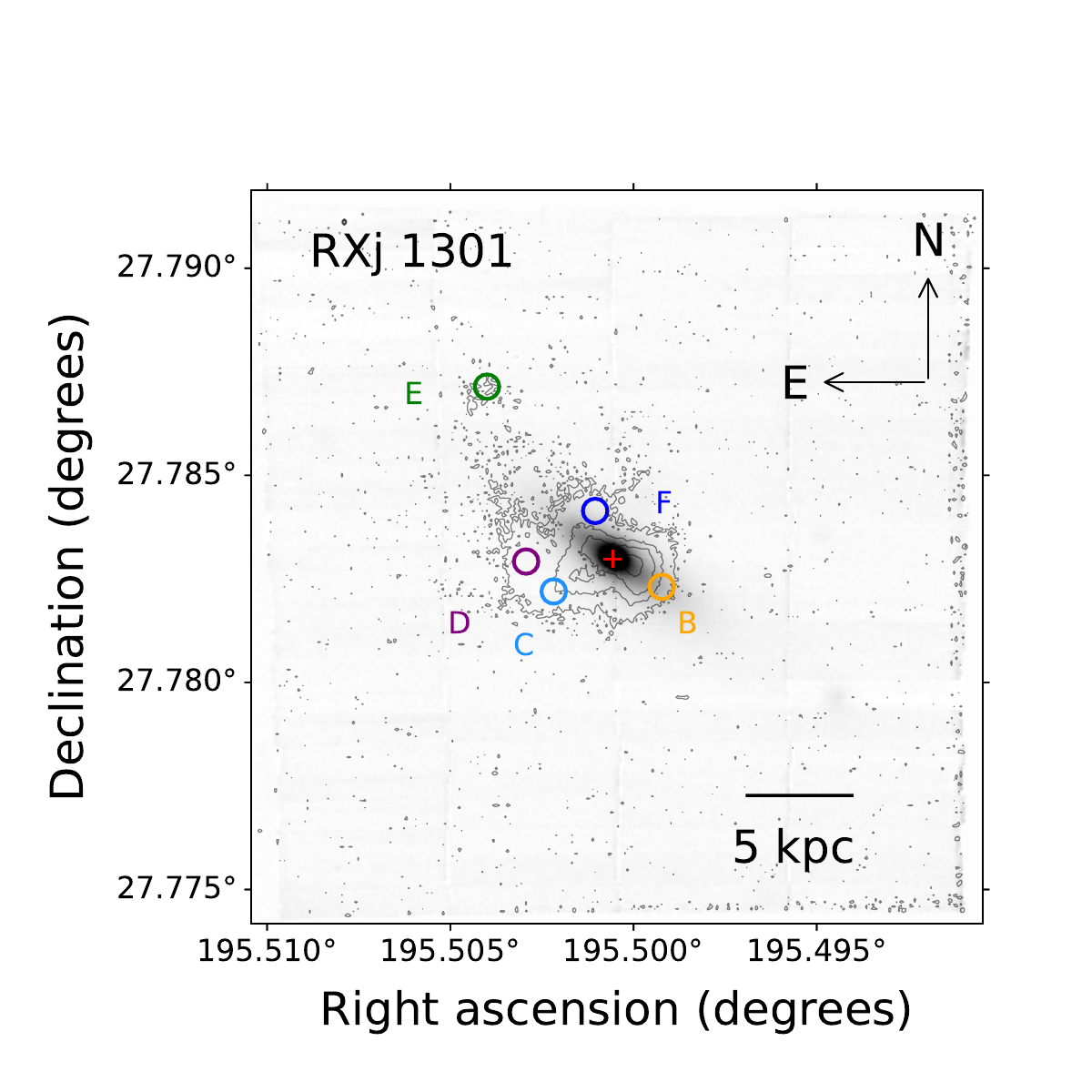}
    \includegraphics[width=0.32\textwidth]{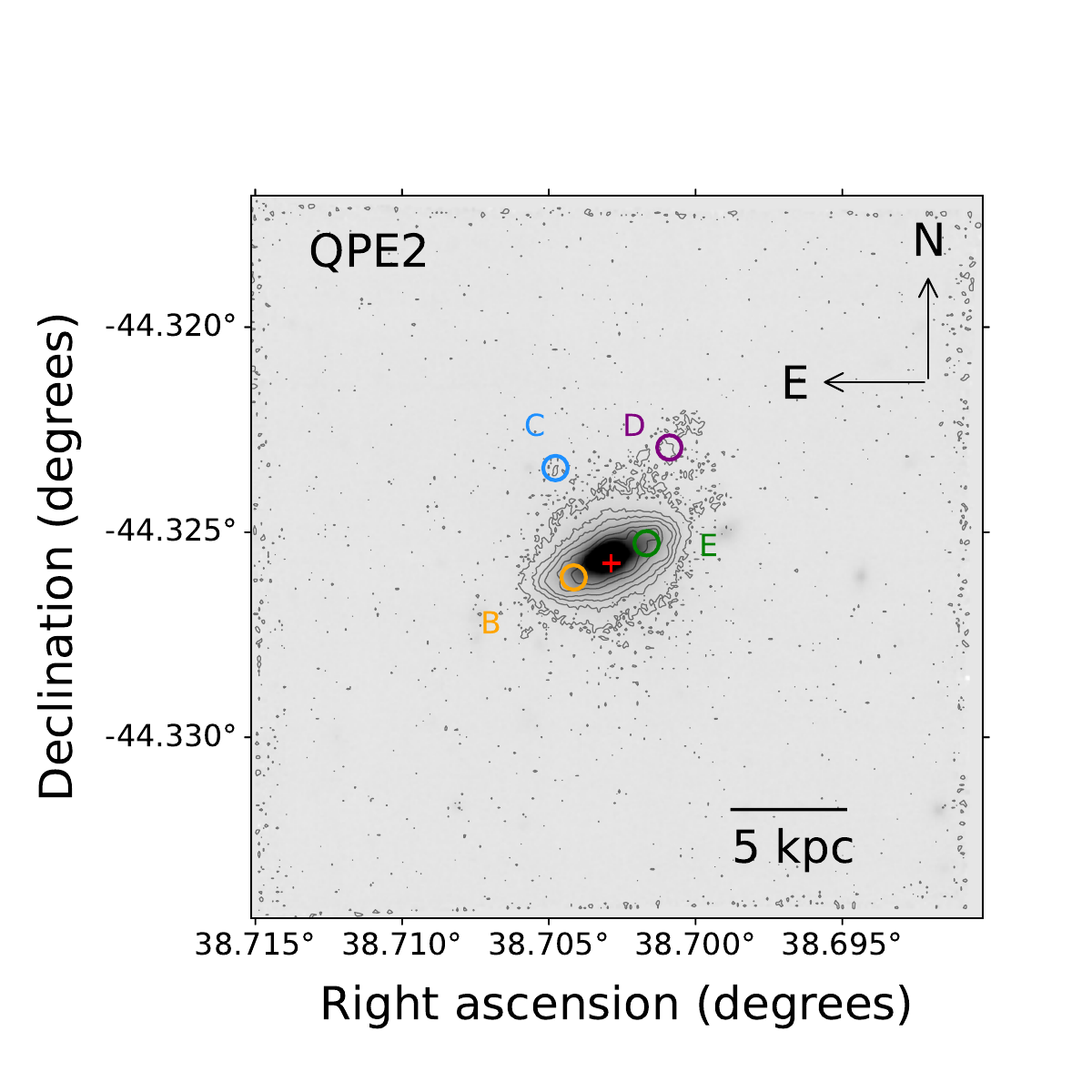}
    \includegraphics[width=0.32\textwidth]{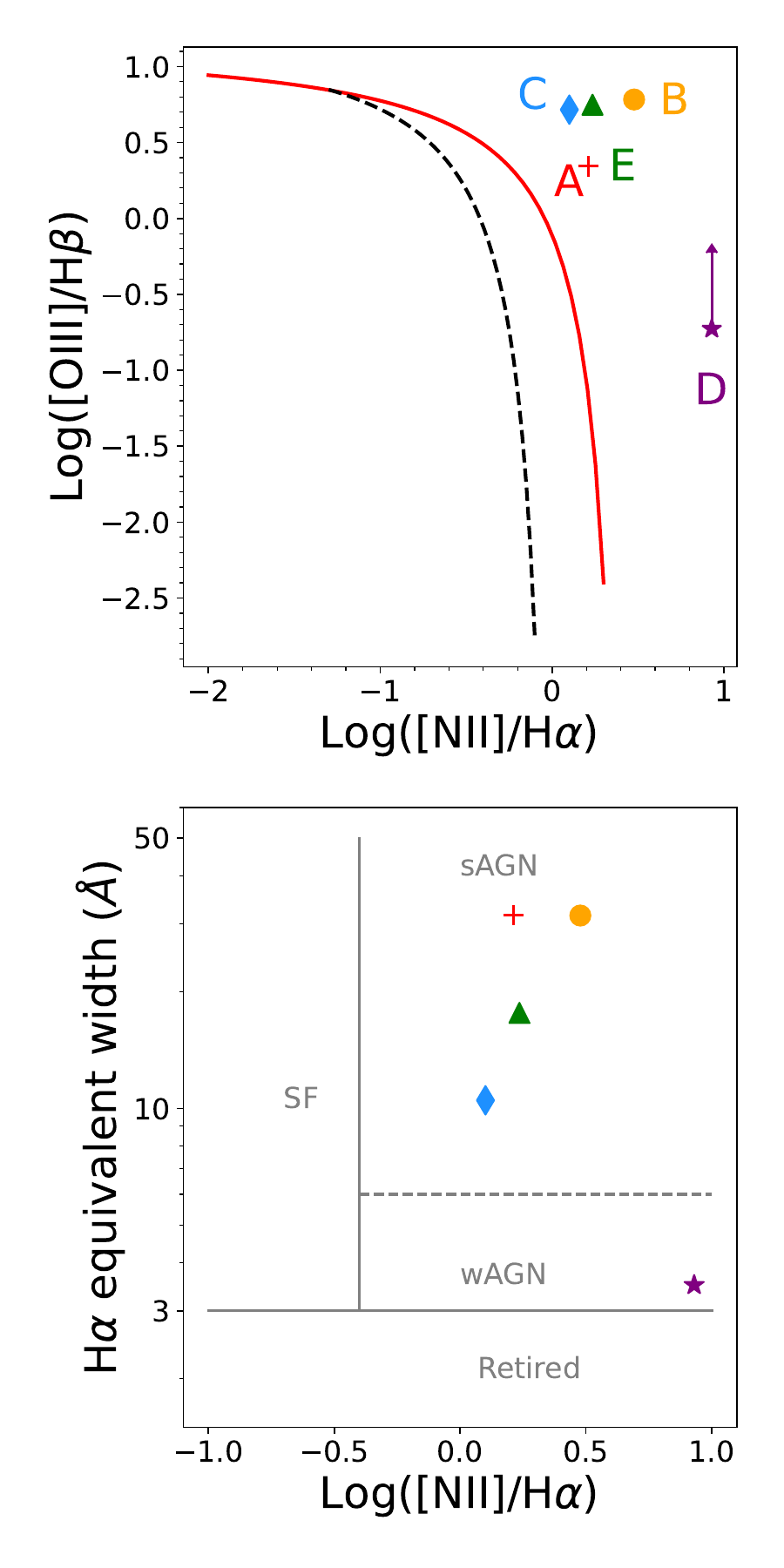}
    \includegraphics[width=0.32\textwidth]{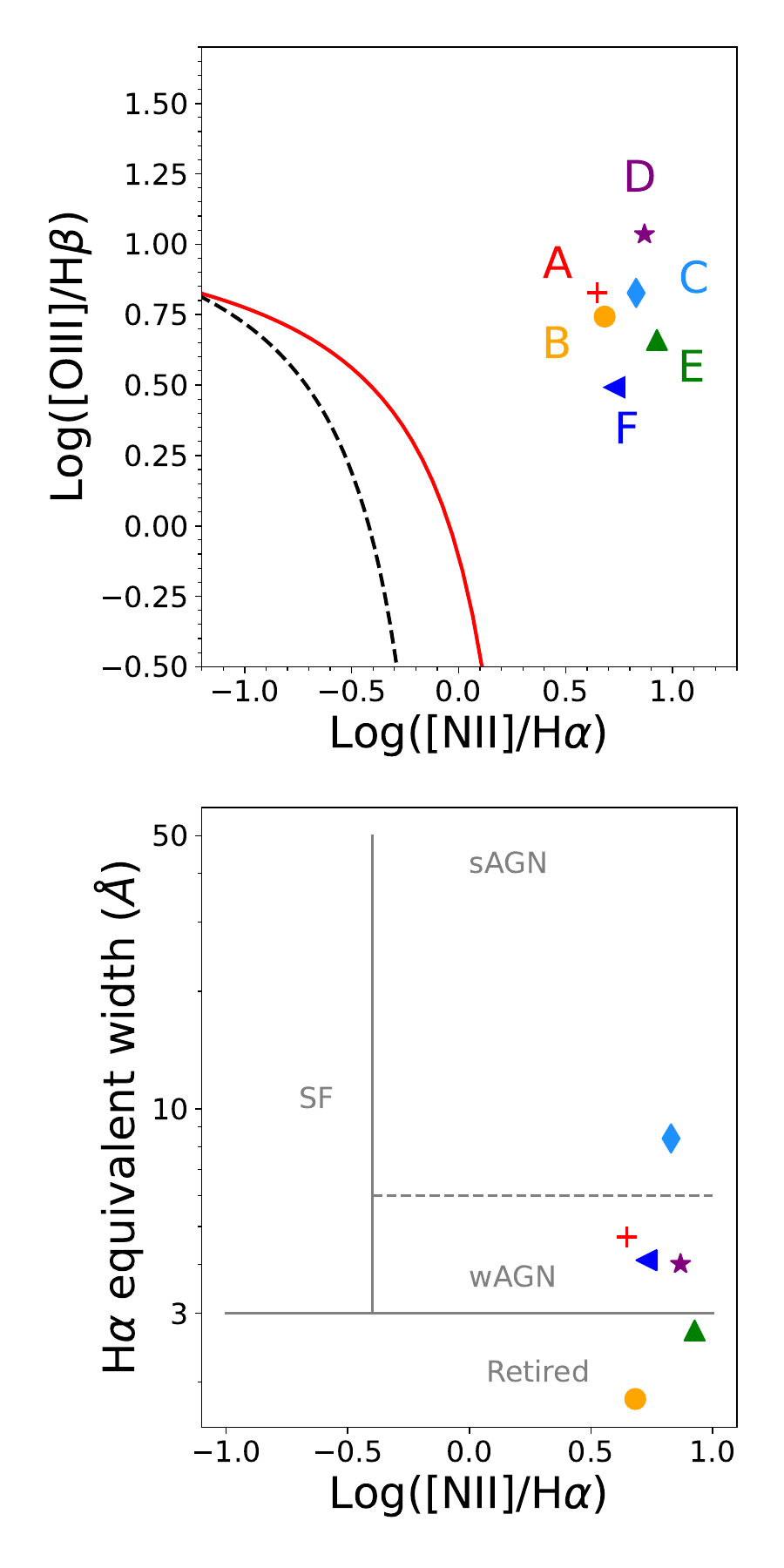}
    \includegraphics[width=0.32\textwidth]{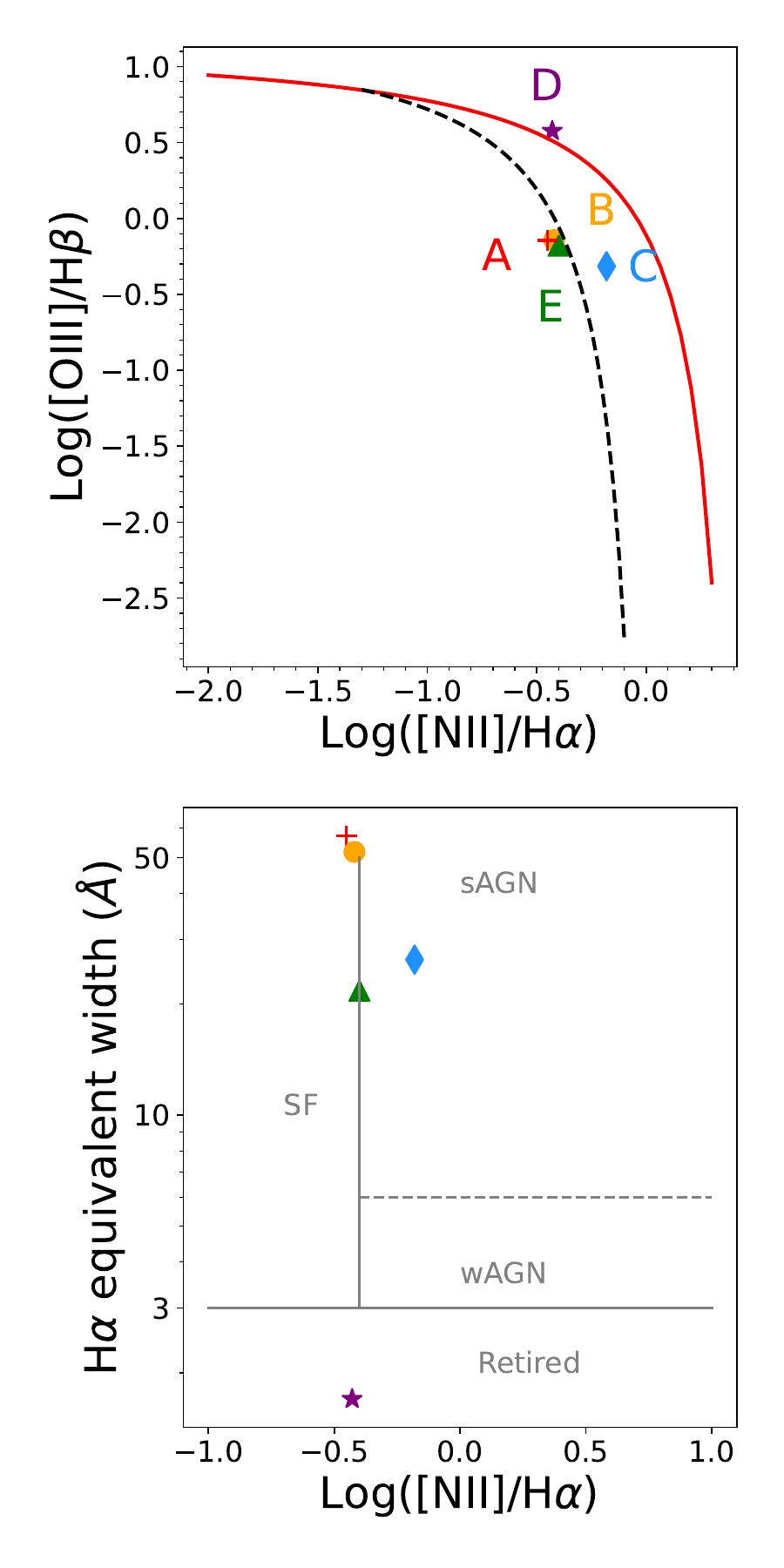}
    \caption{{\bf Images of the host galaxies.} Top: RGB color composites. The continuum is shown in red, while green hues denote H$\alpha$ (QPE2) or N\,\textsc{ii} (RXJ1301 and GSN069). O\,\textsc{iii} is shown in blue. 
    Second row: grayscale continuum images overlaid with a contour map of the emission line. The red cross indicates the continuum nucleus position. Kilo-parsec scale extended emission line regions of ionized gas are discernable in 3/5 QPE hosts. This ionized gas is decoupled from the continuum emission. 
    Third row: BPT diagram of the line ratios using the regions indicated in the panel above, after correcting for the stellar continuum component. Bottom row: WHAN diagram of the same regions.}
    \label{fig:rgb}
\end{figure*}

Upon inspection of the emission line maps (Fig. \ref{fig:rgb}, top row), 3/5 sources show very extended emission line regions (EELRs) that have line ratios consistent with being powered by SF or a non-stellar continuum (Fig. \ref{fig:rgb}, bottom two rows). We note that the sensitivity of our observations is not uniform across the sample, and the non-detections in two sources may be in part due to a sensitivity bias related to a higher redshift for eRO-QPE1 ($z$ = 0.0505, the highest redshift in the sample), and/or a fainter and more compact host galaxy (eRO-QPE1 and eRO-QPE3).

\subsection{Extended emission line region morphology}
\label{sec:eelr}
In Figure \ref{fig:rgb} we show continuum images of the galaxies overlaid with emission line contours of H$\alpha$ or N\,\textsc{ii} (the latter is brighter in the case of GSN069 and RXJ1301). GSN069, eRO-QPE2 and RXJ1301 show EELRs on spatial scales extending up to 10 kpc from the galaxy nucleus. eRO-QPE1 and eRO-QPE3 do not show any extended ionized emission and a compact continuum morphology (Figure \ref{fig:overlay_qpe1}). In the case of GSN069 there are patchy clumps of gas bright only in N\,\textsc{ii} which appear completely isolated to the south and south-east. There is no clear visual connection to the stellar light distribution.
These EELRs do not appear to have a preferred directionality compared to the continuum morphology of their host galaxy, as might be expected if they are AGN-driven outflows or the result of stellar feedback following a strong starburst. These morphologies are instead often (but not exclusively) seen in post-merger galaxies where gas can be deposited at large radii as a result of post-merger splashback \citep{Johnston2008, Schweizer13, Weaver18}. However, there are other mechanisms that can result in extended gas distributions. These include galactic scale outflows powered by previous AGN activity and the accretion/infall of gas from the intergalactic medium (e.g. \citealt{Hafen2019}). 

\subsection{Kinematics}
\begin{figure*}
    \centering
    \includegraphics[width=0.325\textwidth]{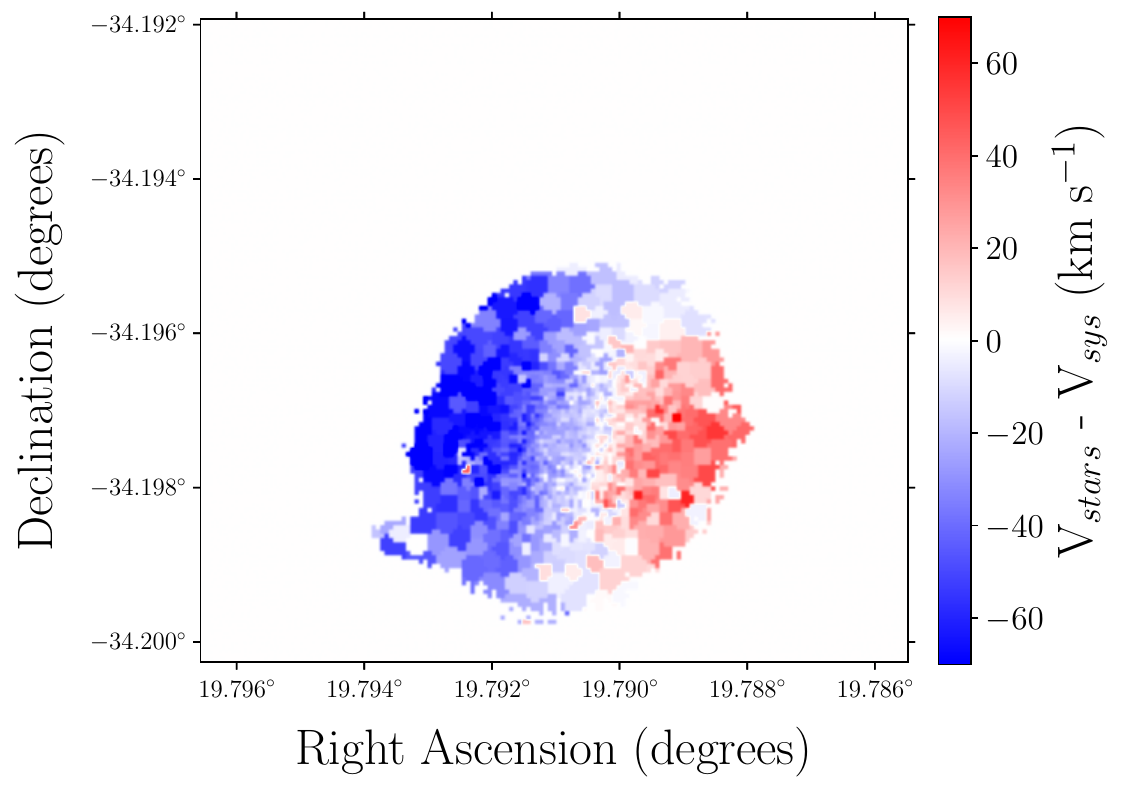}
    \includegraphics[width=0.325\textwidth]{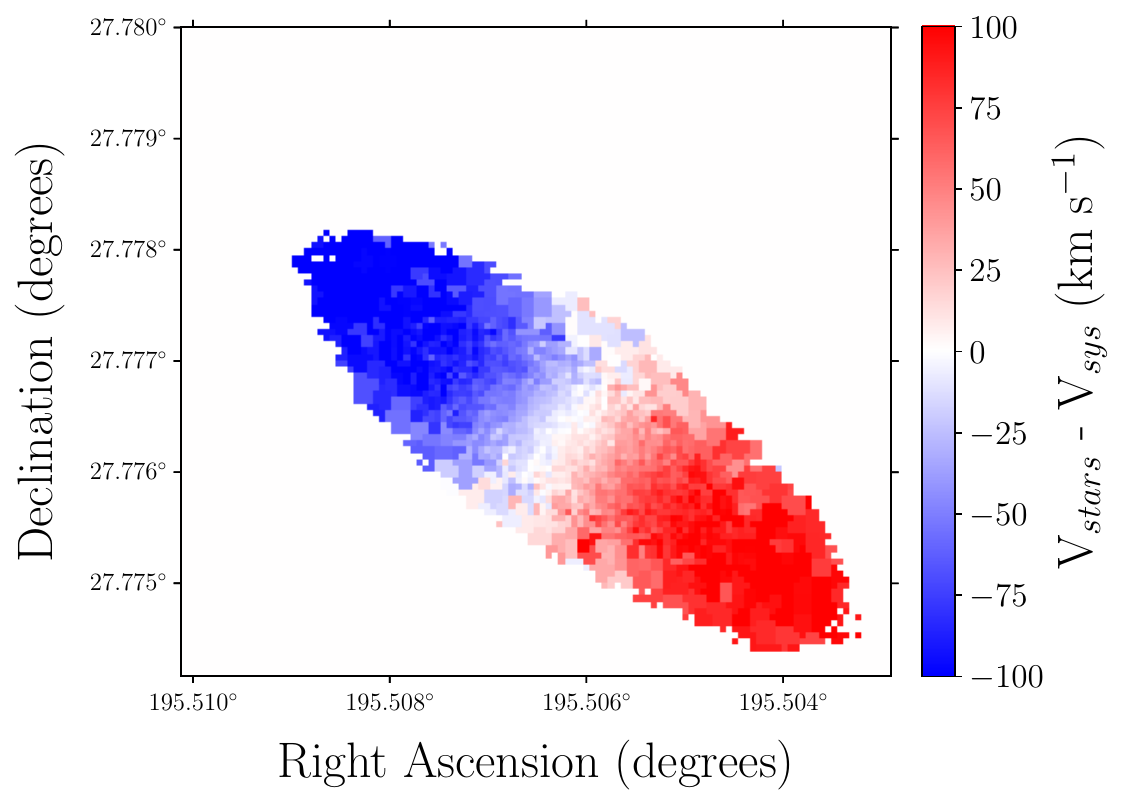}
    \includegraphics[width=0.325\textwidth]{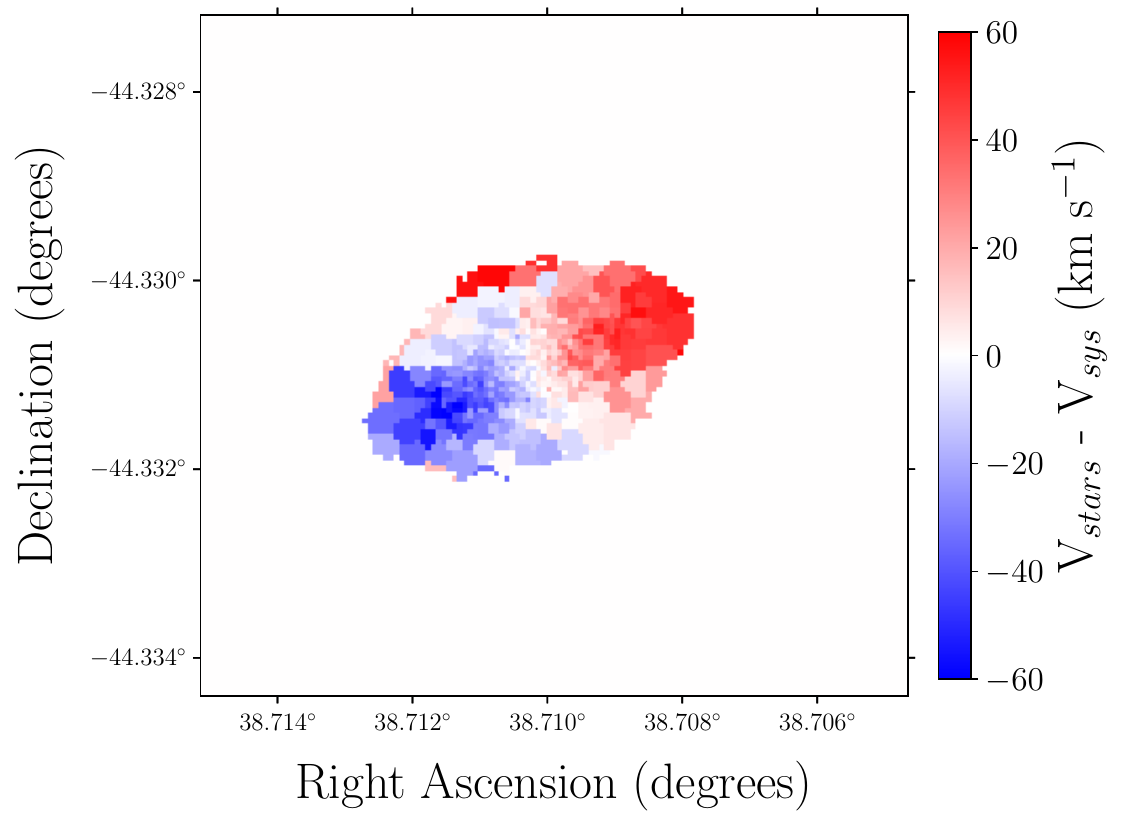}
    \includegraphics[width=0.325\textwidth]{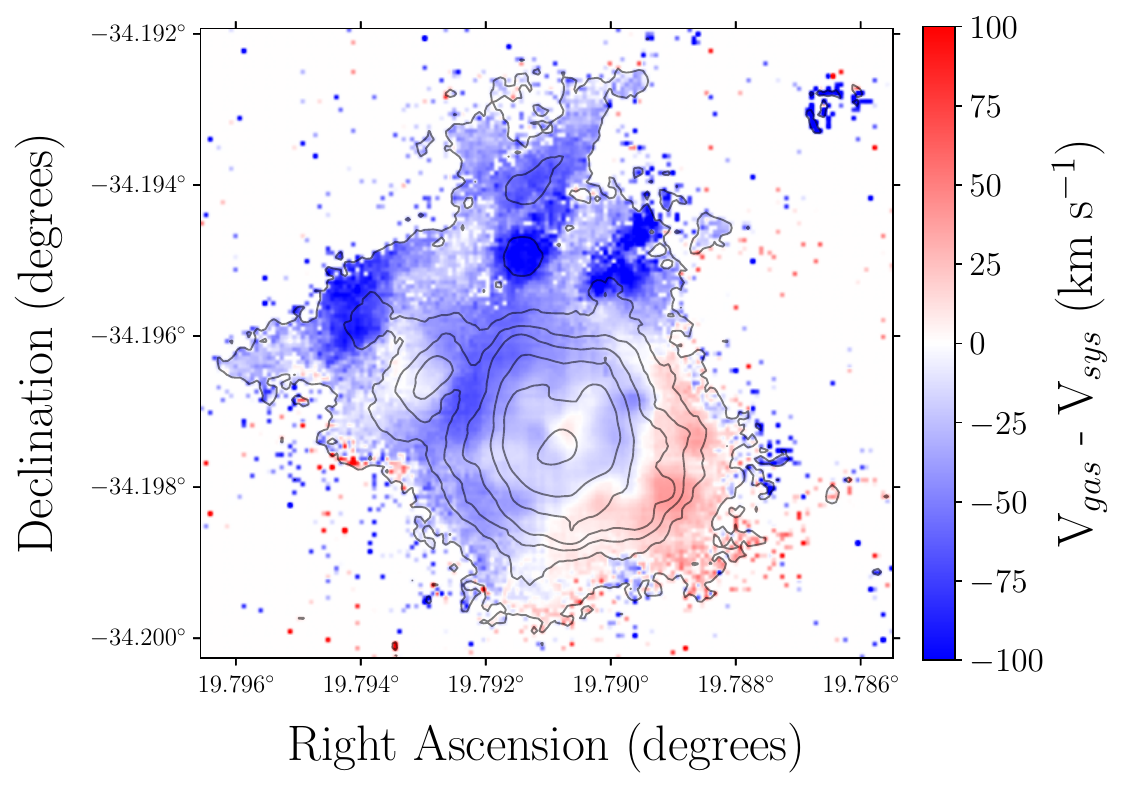}
    \includegraphics[width=0.325\textwidth]{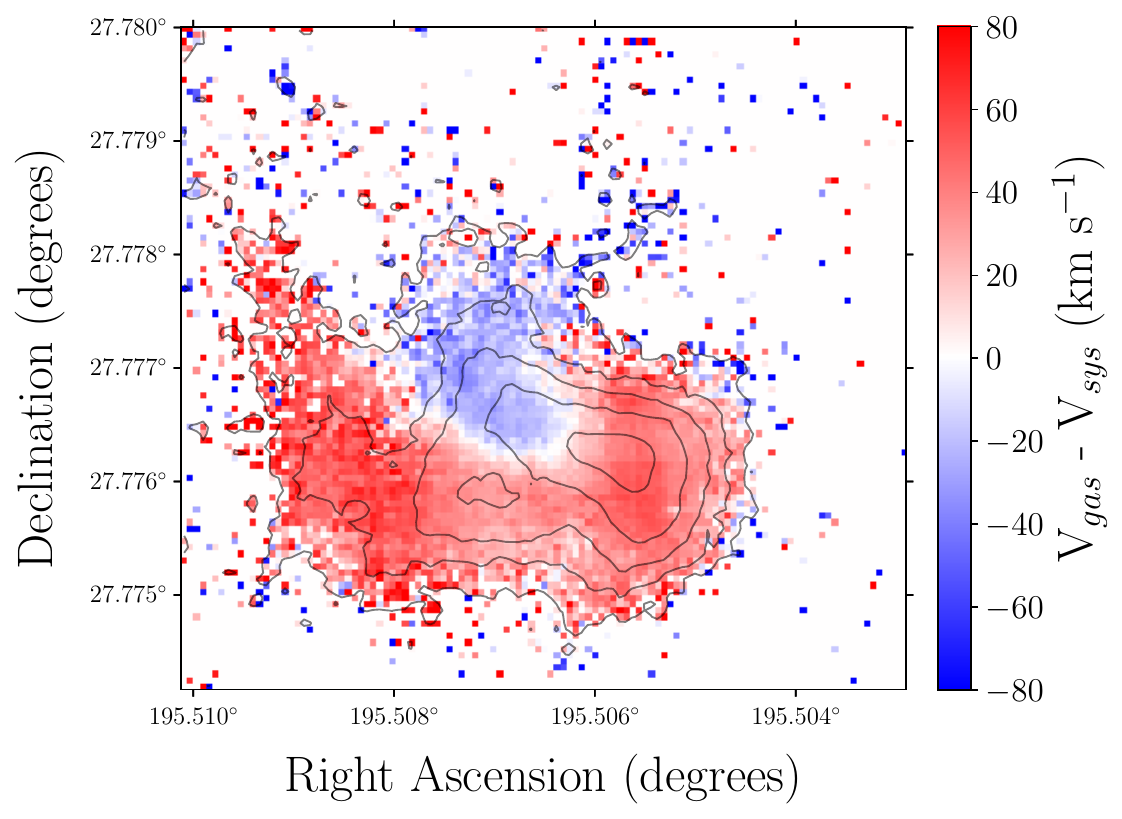}
    \includegraphics[width=0.325\textwidth]{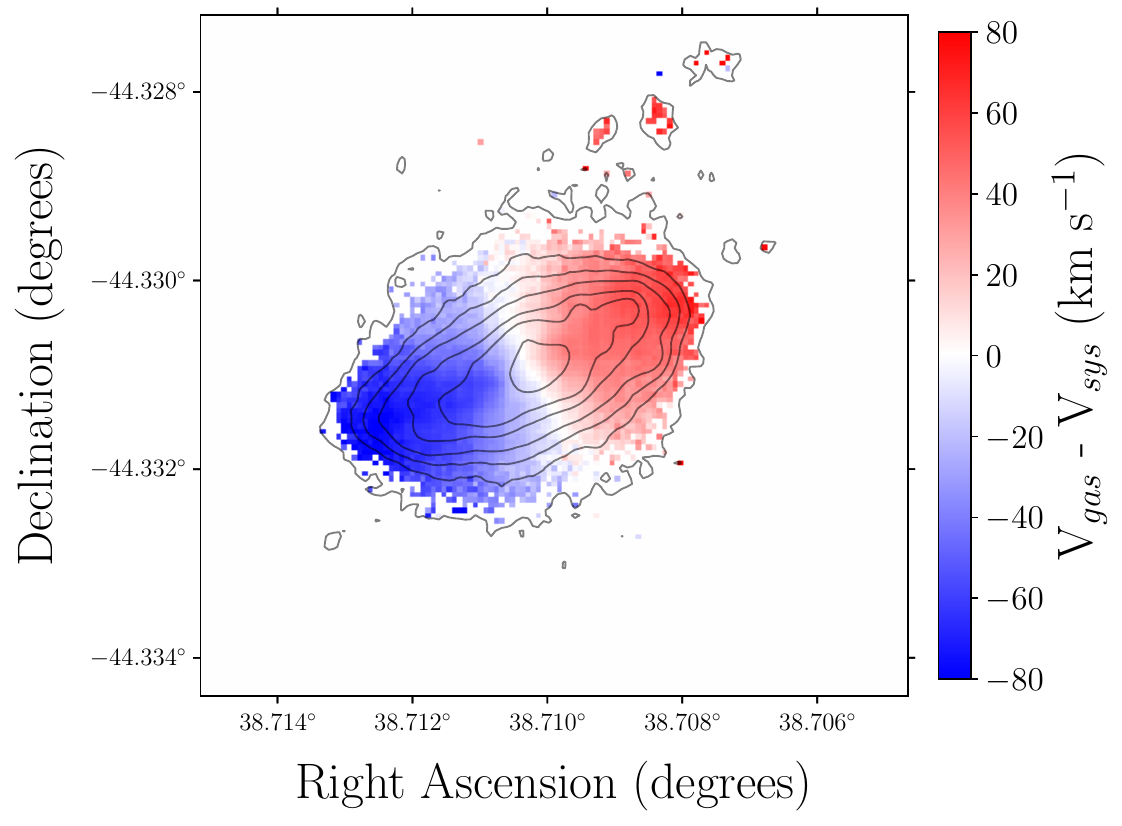}
    \includegraphics[width=0.325\textwidth]{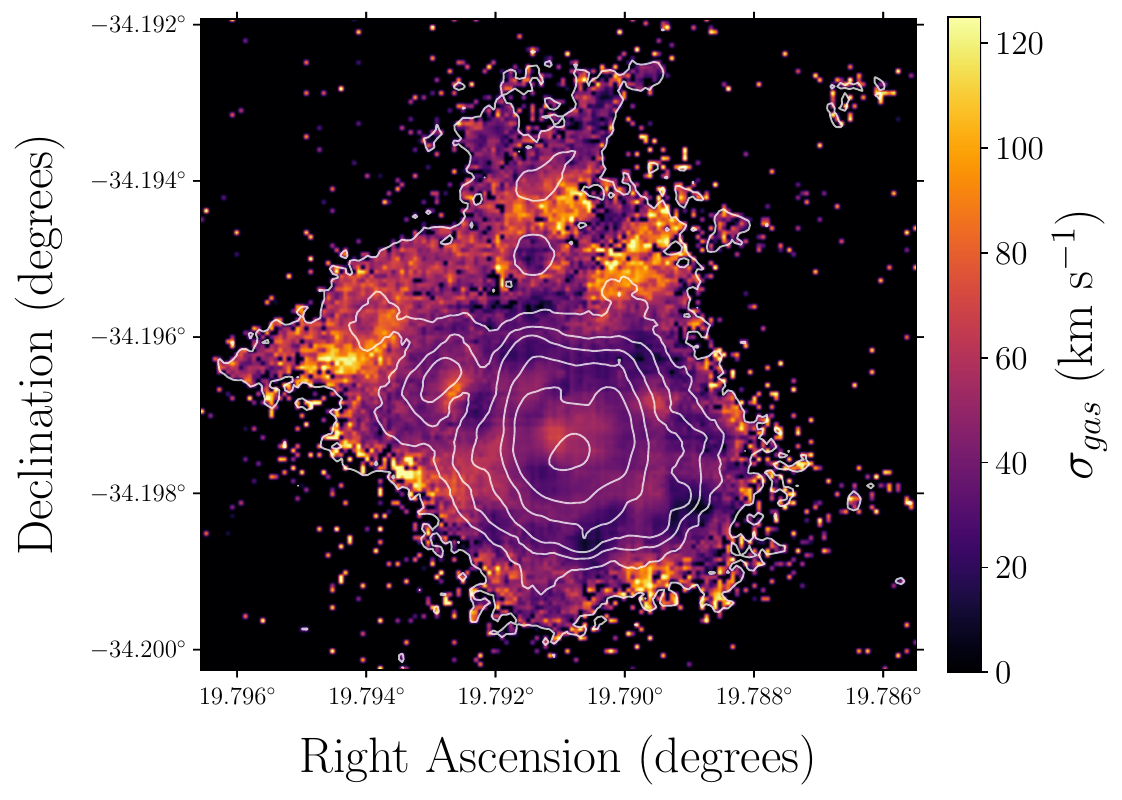}
    \includegraphics[width=0.325\textwidth]{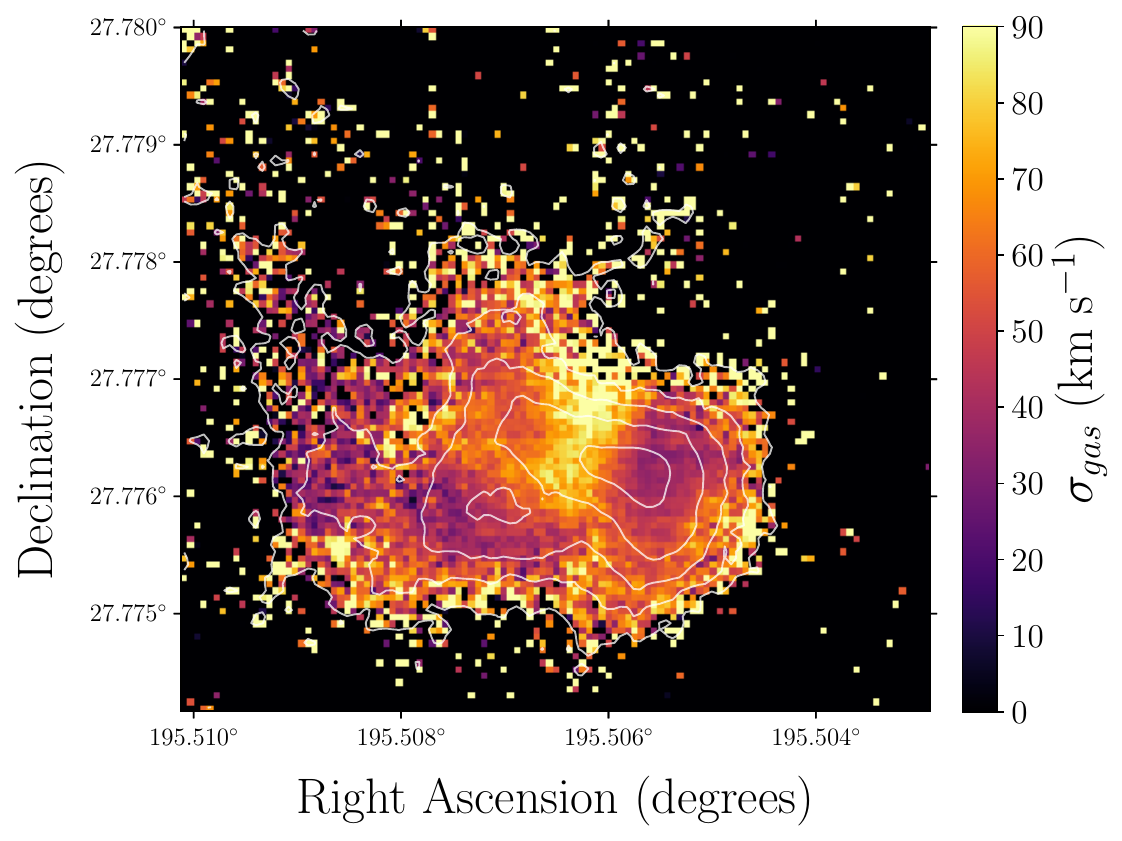}
    \includegraphics[width=0.325\textwidth]{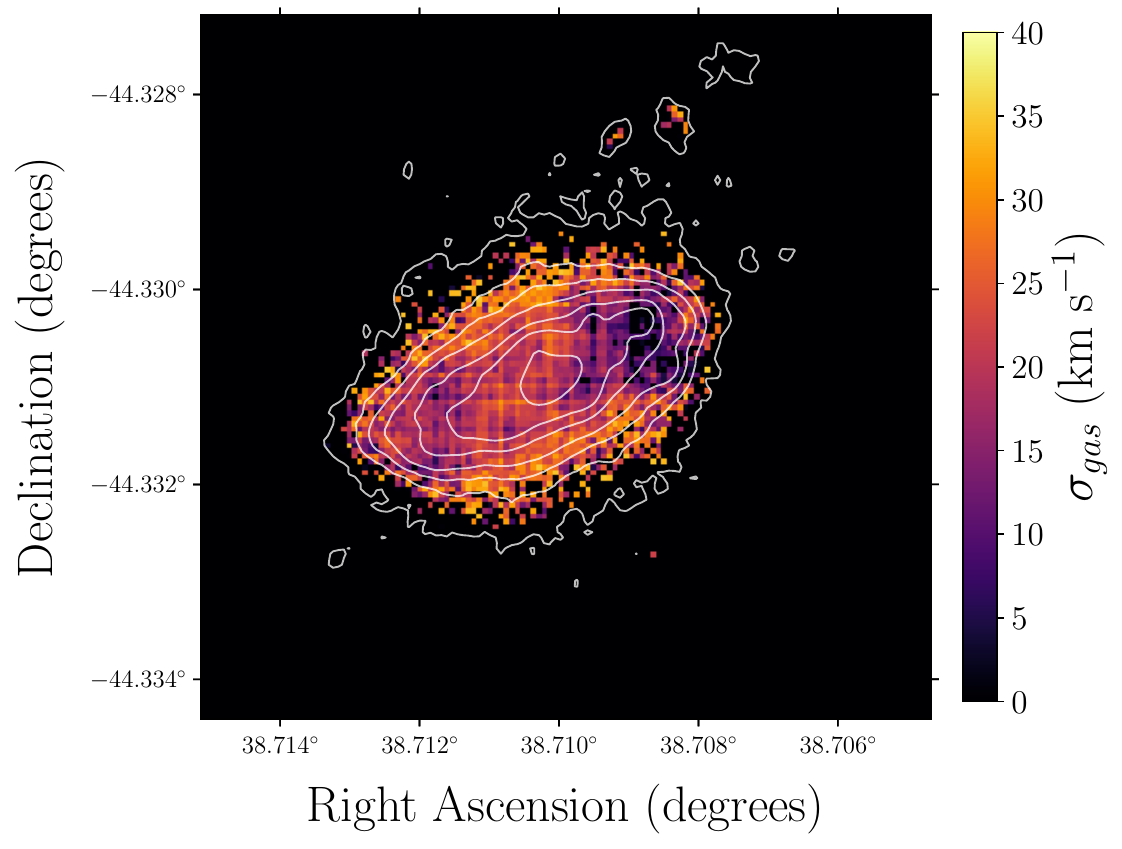}
    \caption{{\bf Spatially resolved kinematics.} Stellar velocity (top), gas velocity (middle) and gas velocity dispersion (bottom) maps. The contours indicate the EELR to guide the eye. The left column shows GSN069, while the middle column shows RXJ1301 and eRO-QPE2 is shown in the right-most column. The sky area is identical for every source across the panels to highlight the differences in EELR kinematics compared to the stellar continuum. }
    \label{fig:kinematics}
\end{figure*}

Spatially resolved kinematics of the stellar and gas components are presented in Figure \ref{fig:kinematics}. The stellar velocity component is well-behaved and displays rotation-dominated profiles in every galaxy (shown in Figure \ref{fig:veldisp} in the Appendix). The central stellar velocity dispersions we measure using aperture spectra are consistent with literature values, except for RXJ1301 and eRO-QPE3. For the former we find a value of 68$\pm$5 km s$^{-1}$, compared to 90$\pm$2 km s$^{-1}$ measured using SDSS data \citep{Wevers22}. This is corroborated by the spatially resolved map, which has peak values around 70 km s$^{-1}$. For the latter source, \citet{Arcodia24} reported a velocity dispersion of 83 km s$^{-1}$, while we find $\sigma_{\rm star} = 38 \pm 5$ km s$^{-1}$. Using the M$_{\rm BH}$--$\sigma$ relation of \citet{Gultekin2009} results in a black hole mass of M$_{\rm BH} = 5.1 \pm 0.55$ M$_{\odot}$ for eRO-QPE3.

The ionized gas kinematics display more disturbed patterns. In GSN069 and eRO-QPE2, the ionized gas does appear to roughly co-rotate with the stellar motions globally, although there is a velocity gradient between the gas and stars in both systems. For RXJ1301, the EELR appears to consist of a single kinematic component with a consistent velocity shift of +50 km s$^{-1}$ (see Fig. \ref{fig:kinematics}, middle panel), completely decoupled from the rotation of the stellar motion. This is very similar to the EELR seen in the TDE host galaxy Mrk 950, presented in a companion paper (Wevers et al. submitted).

More generally, the majority of the ionized gas in these galaxies appears to have very low turbulence, with velocity dispersions that are typically $<$75 km $^{-1}$. We highlight that the gas in eRO-QPE2 has extremely low line widths $<$30 km s$^{-1}$ throughout the galaxy and EELR. In GSN069, there are (disconnected) regions at the edge of the EELR with velocity dispersions up to 150 km s$^{-1}$ (compared to $\sim$75 km s$^{-1}$ in the nucleus) at distances of 7--10 kpc from the centre of the galaxy, although the bulk of the gas has $\sigma_{\rm gas} < 75$ km s$^{-1}$. 

\subsection{Diagnostic diagrams}
We use the emission line ratios of these galaxies in diagnostic diagrams to understand the dominant ionization mechanism. The EELR pseudo-aperture spectra are located within the AGN region of the BPT and WHAN diagrams, shown in the bottom two rows of Figure \ref{fig:rgb}.
Our results for the galactic nuclei are consistent with the conclusion by \citet{Wevers22} that the photo-ionization mechanism in the QPE host galaxies is likely non-stellar in nature, except for eRO-QPE3 whose ionizing continuum appears to be SF-dominated. 
We show the aperture spectra of various regions in Figures \ref{fig:spectra_gsn069}, \ref{fig:spectra_rxj1301} and \ref{fig:spectra_qpe2} in the Appendix.

\subsection{Constraints on the ionizing luminosity}
\begin{figure}
\label{fig:lion}
\centering
    \includegraphics[width=0.5\textwidth]{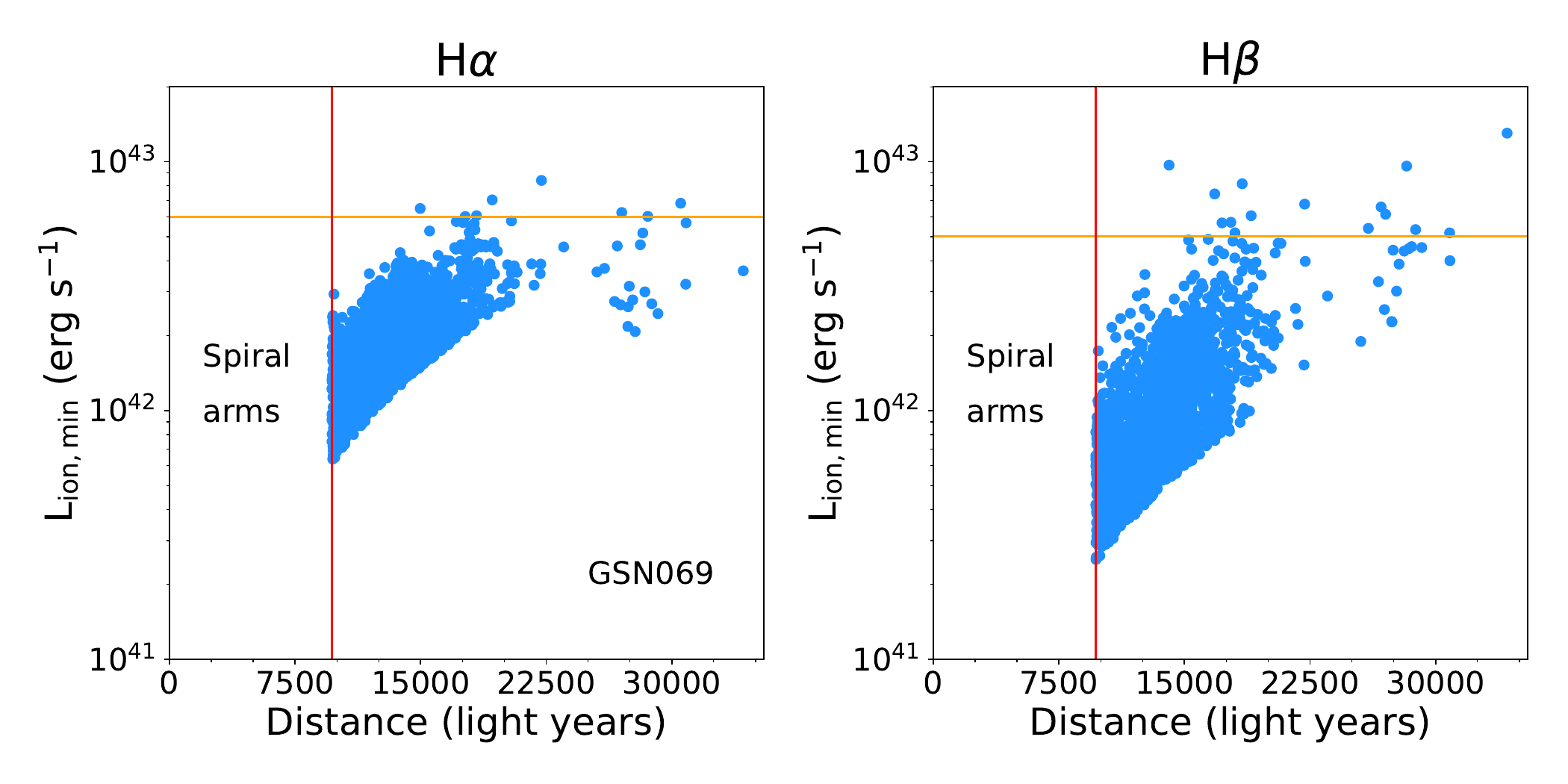}
    \includegraphics[width=0.5\textwidth]{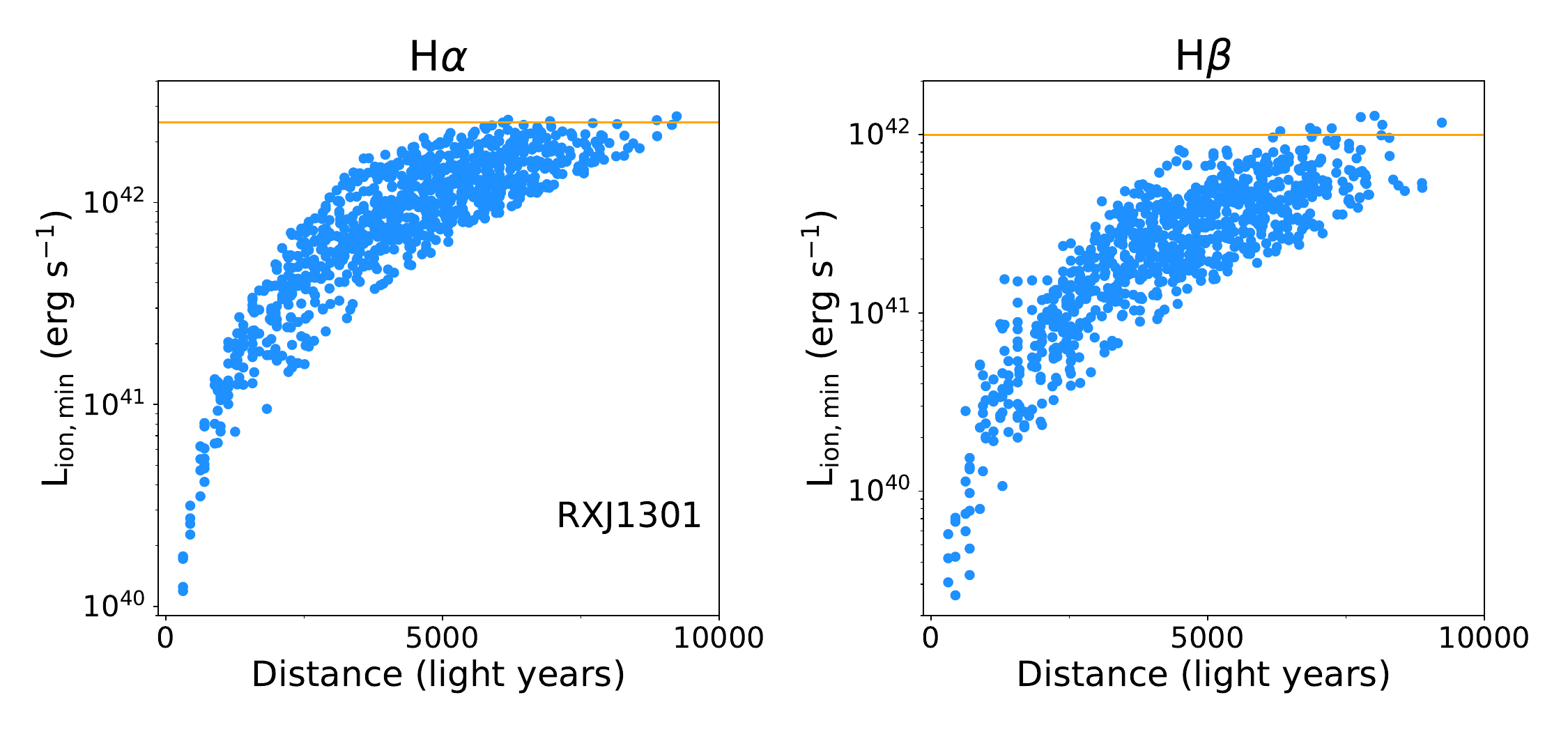}
    \includegraphics[width=0.3\textwidth]{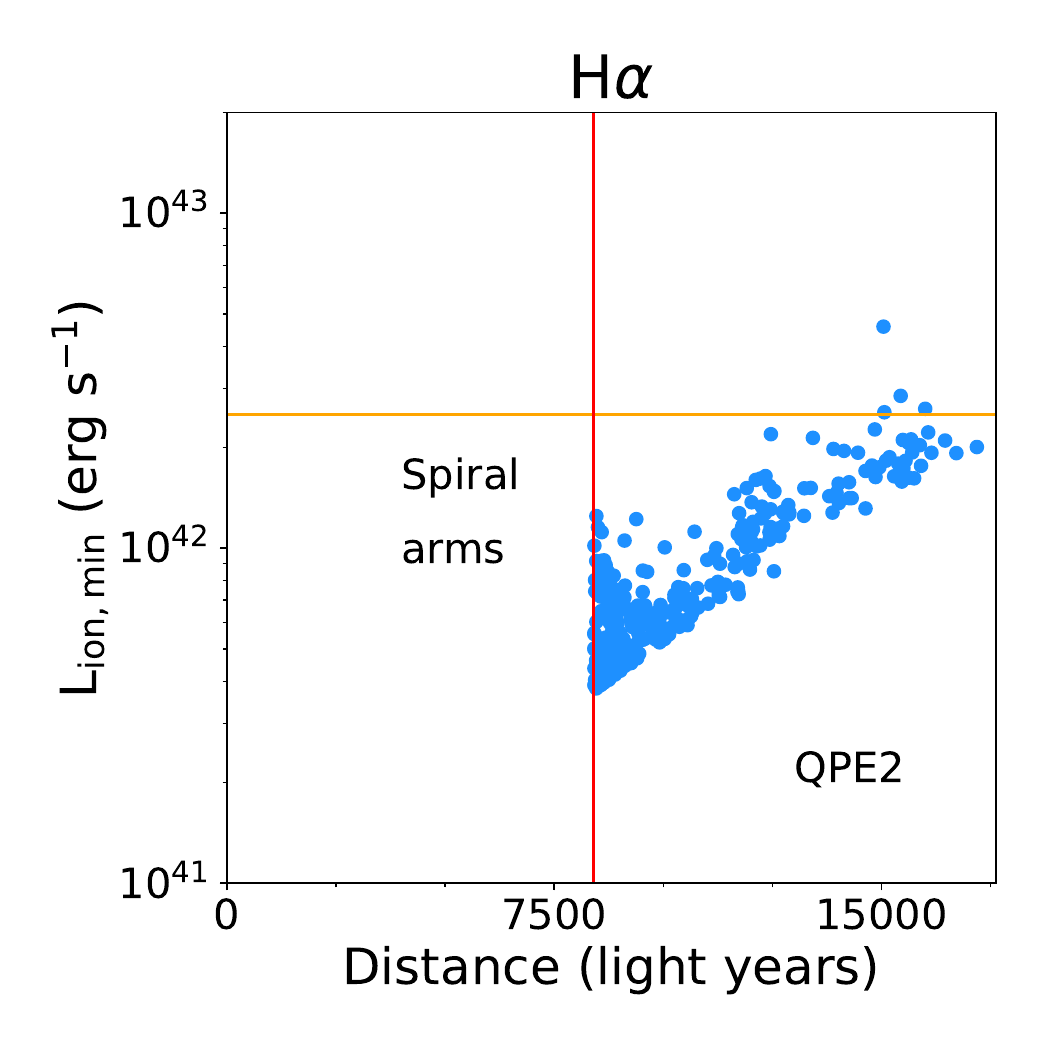}
    \caption{{\bf Constraints on the ionizing luminosity.} Every data point represents the minimum required ionizing luminosity to explain the EELR (indicated by the orange horizontal line), assuming recombination balance, for an individual spaxel. The inner regions of GSN069 and eRO-QPE2 are actively star-forming, so are excluded for clarity (marked by the red vertical lines).}
\end{figure}

The large extent of the EELRs can be used to probe the recent accretion history of the SMBHs in these galaxies. We follow \citet{Keel2017} and \citet{French23} to estimate the required ionizing luminosity to power the observed EELR line fluxes. In the assumption of photo-ionization equilibrium (recombination balance), such an estimate can be made for each spaxel by taking into account the diminishing covering factor (f) with distance (the further away a spaxel is from the nucleus, the higher the required ionizing luminosity for a fixed observed line flux). More quantitatively, we follow \citet{French23} by using the H$\alpha$ and H$\beta$ emission line luminosities $L_H$\footnote{Due to the relatively low SNR of spectra in individual spaxels, we do not incorporate an extinction component to estimate these line fluxes.} to compute two independent estimates of L$_{\rm ion, min}$ in each spaxel (see Fig. \ref{fig:lion}) as follows: 
\begin{equation}
    L_{\rm ion, min} = \frac{n_r L_H E_{\rm ion}}{E_H} \frac{1}{f}
\end{equation}
Here, f is the covering fraction in units of spaxels, 
\begin{equation}
   f = \frac{(2\ \rm arctan(0.5/r))^2}{4 \pi}
\end{equation}
$n_r$ is the number of ionizing photons per recombination, $E_{\rm ion}$ is the ionizing energy between 13.6 and 54.6 eV (we assume that higher energy photons will be primarily absorbed by Helium). $E_H$ is the energy per H$\alpha$ or H$\beta$ photon. 

Following \citet{Keel2012} we further use the Infrared Astronomical Satellite (IRAS) all-sky survey data (derived from the 60$\mu$m and 100$\mu$m measurements) to estimate the present-day AGN luminosity or derive upper limits. A correction for present-day star formation is applied based on the distance from the SF main sequence in the BPT diagram \citep{Wild2010}. These estimates are conservative given the potential for additional extinction not probed by the Balmer decrement (see e.g. \citealt{Baron22}). The results can be found in Table \ref{tab:eelr}. We highlight that for every QPE host with an EELR detection, we infer a present-day nuclear luminosity (L$_{\rm nuc,IR}$) that is below the luminosity required to power the most distant regions of the EELRs. In other words, if an AGN was previously present in these systems it has recently (in the last 1--3 $\times$ 10$^{4}$ yrs) faded in luminosity by a factor of at least 2--5. This is consistent with the episodic nature of AGN activity, which has typical duty cycles of $\sim$10$^5$ years with rapid transitions on $\sim$10$^4$ year timescales \citep{Keel2017, Schawinski15}.

\begin{table*}
    \caption{Overview of the QPE and EELR properties of the host galaxies. Class indicates the spectroscopic classification based on the H$\delta_A$ absorption and H$\alpha$ EW: post-starburst (PSB), quiescent Balmer strong (QBS) or neither (N). Burst age and fraction refer to the time since the most recent starburst and the percentage of stars formed in said starburst (see text), where numbers in brackets denote uncertainty on the last digit. Ion. cont. indicates the classification of the ionizing continuum, based on the BPT and WHAN diagnostic diagrams. R$_{\rm max}$ refers to the largest (projected) extent of the EELR relative to the nucleus. L$_{\rm QPE,quiesc}$ is the bolometric luminosity during QPE quiescence, taken from the literature. L$_{\rm ion, min}$ is the minimum nuclear ionizing luminosity required to power the EELR. L$_{\rm IR}$ is the luminosity inferred from IRAS FIR detections or upper limits.}
    \centering
    \begin{tabular}{c|cccccccccc}
    Source & Class & Burst age & Burst frac. & EELR & Ion. cont. &  R$_{\rm max}$ & L$_{\rm QPE,quiesc}$ & L$_{\rm ion, min}$ & L$_{\rm nuc,IR}$  \\
     & & log$_{10}$(yr) & &  &  (EELR) & (lightyr) & (erg s$^{-1}$) & (erg s$^{-1}$) & (erg s$^{-1}$) \\\hline
        GSN069   & N & Ongoing & 0.045(2) & Y &  Non-stellar & 30\,000 & 3$\pm$2$\times$10$^{42}$ & $>$6$\times$10$^{42}$ & $<$3$\times$10$^{42}$ & \\
        RXJ1301  & PSB & 8.68(1) & 0.12(1) & Y &  Non-stellar & 9\,000 & 2.5$\pm$0.5$\times$10$^{42}$ & $>$2.5$\times$10$^{42}$ & $<$1.5$\times$10$^{42}$ \\
        eRO-QPE1 & QBS & 9.11(4) & 0.066(2) & N &  --- & --- & 5.1$^{+1.6}_{-1.5} \times$10$^{40}$ & --- & --- &  \\
        eRO-QPE2 & N & Ongoing & 0.14(2) & Y & Composite & 15\,000 & 3$^{+4}_{-2} \times 10^{43}$ & $>$2.5$\times$10$^{42}$ & 4$\times$10$^{41}$ \\
        eRO-QPE3 & N & 8.6(2) & 0.014(1) & N & --- & --- & $<$4$\times$10$^{40}$ & --- & --- & \\
    \end{tabular}
    \label{tab:eelr}
\end{table*}

\section{Discussion}
\label{sec:discussion}
Our analysis shows that 4/5 QPE host galaxies  have one or more peculiar properties, including the presence of a quenched starburst (a post-starburst phase, RXJ1301), a quiescent Balmer strong spectrum (which may indicate a weaker recent starburst, eRO-QPE1), the presence of extended ionized gas emission (Fig. \ref{fig:kinematics}, RXJ1301, GSN 069 and eRO-QPE2) that is decoupled from the stellar continuum, and/or the presence of a recently faded nuclear engine in three systems (Table \ref{tab:eelr}, RXJ1301, GSN 069 and eRO-QPE2). The latter occurs in $\sim$1/3 of EELR hosting galaxies \citep{Keel2012}, so the unity fraction observed here may be another peculiar characteristic of QPE hosts but requires confirmation with larger samples. No tidal tails or other visual disturbances are obvious in the stellar continuum light. Such features are typically seen in other studies of Voorwerp / faded AGNs with EELRs (e.g. \citealt{Keel2015}), and they provide a robust association for a post-merger origin of the EELR gas. We note that at present we do not have similarly compelling evidence for a post-merger origin because our observations are relatively shallow and have poor spatial resolution. Further studies of larger samples are needed to confirm the tentative connection between the QPE host properties and post-merger faded AGNs with EELRs.

We find a fractional incidence of EELRs in QPE host galaxies of f$_{\rm EELR}$ = 0.6$^{+0.4}_{-0.33}$ (where the uncertainties are two-sided Poisson confidence intervals, \citealt{Gehrels1986}). 

We compare this to literature values with the caveat that the galaxy samples are not matched in properties (e.g., redshift, stellar mass, depth of observation) because only a few statistical studies exist. The QPE host EELR incidence is a factor of $\sim$10 higher than the fraction of EELRs found in the general galaxy population, f$_{\rm EELR} \sim 0.085$, although a large number of these are SF- or AGN-driven outflows \citep{LopezCoba2020} which is not the case for the EELRs discovered in this work. The incidence of EELRs in QPE hosts is a factor of $\sim$30 higher than in AGN host galaxies (f$_{\rm EELR}$ = 0.023$\pm$0.013, \citealt{Keel24}). 

Compared specifically to PSB (E+A) galaxies, \citet{French23} found the EELR incidence to be f$_{\rm EELR}$ = 0.065$^{+0.035}_{-0.025}$, and hence we infer an overrepresentation by a factor of $\sim$10$\times$ even compared to this rare type of post-merger galaxy comprising less than 0.2 per cent of the local galaxy population \citep{Zabludoff1996}.
The EELR fraction of QPE hosts is similar to that seen in post-merger galaxies identified through the presence of tidal tails and interaction with companion galaxies in continuum light \citep{Keel24}. The lack of similar tidal features may indicate that the QPE hosts are not post-merger (Section \ref{sec:eelr}), or that the merger occurred $\sim$1 Gyr ago, short enough that we still see the post-merger signatures such as extended reservoirs of gas but long enough that
most obvious merger features (e.g. dust lanes, tidal tails) have already faded \citep{Pawlik16}.

With the caveat of small sample statistics, we note that the overrepresentation of EELRs in QPE hosts is somewhat higher than that found in TDE host galaxies (f$_{\rm EELR}$ = 0.19$^{+0.14}_{-0.11}$ for the entire sample with MUSE observations, Wevers et al. submitted) but consistent with the PSB-TDE-EELR host galaxies (f$_{\rm EELR}$ = 0.6$^{+0.4}_{-0.33}$, Wevers et al. submitted)\footnote{We emphasize that the sensitivity of the IFU observations is not uniform across the QPE and TDE samples.}. The implication is that there may be a connection between these two classes of nuclear transients, which we explore further below. 

\subsection{The QPE--TDE host galaxy connection}
In a companion paper (Wevers et al. submitted) we analyzed MUSE observations of 16 TDE host galaxies. Combined with previously known EELRs in TDE hosts, we inferred an elevated EELR incidence compared to post-starburst galaxies as well as the general galaxy population. The overabundance of EELRs among TDE hosts is a factor of 10$\times$ compared to post-starburst galaxies, which is similar to the overrepresentation reported for QPE hosts in this work. 

In addition, the QPE host galaxy EELRs show peculiar kinematics that are strikingly similar to those of the EELR-hosting TDE host galaxies. The combination of non-stellar photo-ionization, a recently faded AGN, low velocities and low velocity dispersions is rare among the general galaxy population ($<$ 1 per cent), although this is based on HST narrow-band imaging rather than IFU observations \citep{Keel2012, Keel24}. These properties are inconsistent with AGN- or shock-driven kinematics and associated ionization mechanism. 

A tentative connection between the QPE phenomenon and TDEs has been proposed based on the observational properties of QPEs, including long-term declines in the continuum X-ray lightcurves \citep{Miniutti19, Miniutti23, Arcodia24}, the presence of a single QPE-like flare in the decay of some TDE candidates \citep{Chakraborty21, Quintin23} and the PSB/QBS preference of the host galaxies \citep{Wevers22}. 
Moreover, several theoretical models invoke partial TDEs as a mechanism to render a (pre-existing) closely bound stellar remnant visible as it interacts with the newly formed, compact accretion disk. Such models strongly disfavor the presence of a large AGN accretion disk because ablation of the star would severely shorten the lifetime of the QPE source \citep{Linial24}.

The tentative connection between TDEs and QPEs can also be investigated using sample properties as is done here.
Our results provide the strongest, albeit indirect, indication to date that QPE and TDE rates are enhanced in gas-rich environments with faded AGNs, and possibly in post-merger systems\footnote{For the QPEs, this statement relies on a single PSB galaxy, RXJ1301.}. The implication could be that they share a common formation path, although the implications of this host galaxy connection depend on the theoretical scenario invoked to explain QPEs.

\subsection{The QPE--EELR connection}
Our analysis shows that the QPE quiescent luminosity can in principle reach the threshold required to power the EELR energy budget (i.e. L$_{QPE,quiesc} \gtrsim L_{\rm ion,min}$) in RXJ1301, eRO-QPE2, but not in GSN069\footnote{Note that the long-term luminosity behavior of this source is highly variable and was significantly higher in the past (with L$_{\rm QPE,quiesc} > $L$_{\rm ion,min}$).} (see Table \ref{tab:eelr}). 
The quiescent luminosity of the 2 host galaxies without EELRs are the lowest among the sample. 
More generally, with the low duty cycle of QPEs ($\sim$25\%, \citealt{Guolo2023}) and their unknown but likely limited lifetime\footnote{In the star-disk interaction models, it is $<$1000 yrs \citep{Linial24}.} it appears improbable that the eruptions alone can power the EELRs. 
We note, however, that some scenarios require a (partial) TDE to occur before QPEs can be detected. 
As TDEs typically have a larger energy budget available, it is in principle possible to power the EELRs through a (time-limited) increase in the TDE rate by a factor of $\sim$10 compared to the average TDE rate in PSB galaxies (Wevers et al. submitted). Depending on the relative rate of TDEs and QPEs, it may therefore also be possible that the EELRs are powered (in part) by episodic accretion in the form of an elevated TDE rate, rather than a classical AGN phase. 

\subsection{Implications for the nature of QPEs}
\subsubsection{Partial TDE scenario}
The fact that both QPEs and TDEs prefer host galaxies with similar, peculiar properties can point to an intrinsic connection between these classes of events. In this case, the most obvious link may be through the disturbed state of the galactic nuclei following a (minor) galaxy merger. This would perturb the orbits of stars in the nucleus and lead to an increased rate of star-SMBH encounters of all kinds, including full tidal disruptions, partial TDEs and even direct captures. QPEs have been linked to partial TDEs, where the Hills mechanism can disrupt a binary star system to capture one of the components on a tight eccentric orbit, leading to the repeated partial stripping at each pericentre passage or alternatively direct mass transfer (e.g. \citealt{King22, Krolik22, Linial23a, Lu22, Metzger22, Zhao22}). In this scenario, repeating nuclear transients on all timescales would likely share a common formation channel and therefore host galaxy preference, even though they may not share the same observational properties (e.g. \citealt{Guolo2023}). This idea can be tested through a host galaxy study of a more inclusive set of repeating nuclear transients, including QPEs and partial TDEs recurring on timescales of 10s/1000s of days (e.g. \citealt{Payne21, Wevers23, Liu23, Guolo2023}). 

\subsubsection{Star-disk / BH-disk interaction scenario}
In addition to the partial TDE scenario, there exist related but distinct theoretical models to explain QPEs. For example, the rate and lifetime of EMRIs is sufficiently high that a pre-existing bound stellar-mass companion is expected to be present in numerous galactic nuclei \citep{Linial24}. Rendering such systems visible is possible if a TDE provides a compact accretion disk with which the bound remnant periodically interacts, leading to the QPE phenomenon \citep{Linial23}. In this scenario there is a natural commonality between the host galaxies of QPEs and TDEs, as the latter are required to observe the former. One distinction with the pTDE models is that the typical lifetime of the QPE (set by the ablation timescale of the star while interacting with the disk) is at most 1000 years, and the implication is that the QPEs themselves cannot be invoked as the powering source for the EELR (because their extent is $>$10$^4$ yrs). A preference for post-starburst galaxies was predicted in this scenario, as such systems are more likely to host EMRIs in the first place \citep{Metzger22}.

\subsubsection{Accretion disk instability scenario}
Alternatively, QPEs have also been suggested to be caused by instabilities in large AGN-like disks (e.g. \citealt{Sniegowska20, Raj21, Pan23}). On one hand, it seems less likely that the connection between QPEs and TDEs is intrinsic in this case. The TDEs found in EELR-hosting galaxies are unremarkable in the context of the known TDE sample, while it has been predicted that AGN disks may significantly alter the observed properties \citep{Chan2019}. On the other hand, it has been proposed that the presence of an AGN disk may significantly increase the TDE rate (e.g. \citealt{Kennedy16, Wang2024, Kaur24}). All three QPE nuclei with EELRs have faded in the recent past, which has been suggested to result in a higher TDE rate due to star-disk interactions and the presence of stars formed in the outer regions of the accretion disk \citep{Wang2024}, or alternatively through an enhanced loss cone filling rate through collisionless processes in gas-rich nuclei \citep{Kaur24}. 

We highlight that based on X-ray spectral modeling, the neutral (and ionizing) absorbing columns are very low: $N_{\rm H} \simeq 4\times 10^{21}$~cm$^{-2}$ [E(B-V)$\sim$0.6, \citealt{Guver09}] for eRO-QPE2, and $N_{\rm H} < 5\times 10^{20}$~cm$^{-2}$ [E(B-V)$<$0.07] for all others, i.e., QPE nuclei are completely unobscured. Nevertheless, no broad emission lines are detectable in deep long-slit data \citep{Wevers22} nor in our MUSE observations. For the source GSN 069, the present-day continuum (quiescent) bolometric luminosity is of order few $\times$ 10$^{43}$ erg s$^{-1}$, and hence this lack of broad emission lines (BELs) is unlikely to be the consequence of a too low nuclear luminosity. Instead it may be related to the presence of a disk that is too compact to support a typical broad line region (BLR), which is compatible with the accretion disks that form in the aftermath of a TDE (for which several lines of evidence exist in GSN 069, \citealt{Sheng21, Miniutti23, Patra24}).
\citet{Pan23} argue that while QPEs do require some fine-tuning (e.g., a very small unstable region, peculiar magnetic field, etc.), it is possible, in principle, that these conditions are only met in compact TDE disks but inefficient in much larger AGN disks.

The current weak state of the nuclei (compared to a more luminous state in the past, required by the EELR energy budget) may be the result of a selection bias against QPEs in higher luminosity AGN. It may be more difficult to detect the QPEs as an additional component on top of the AGN continuum emission \citep{Miniutti23} which may be highly variable, especially so for low black hole masses (e.g. \citealt{Ponti12}). Similarly, the TDE detection rate may be biased against AGN host galaxies because of spectroscopic selection biases. In this scenario the selection biases in both TDEs and QPEs would need to {\it conspire} to create the observed similarities and preferences in an extremely rare type of host galaxy. This scenario is somewhat contrived, but ruling it out completely will only be possible with larger samples of QPE and TDE host galaxies. 

\subsection{Localization of low frequency gravitational wave sources}
Maximizing the scientific value of GW sources will be facilitated by complementary EM information such as the host redshift, galaxy/black hole mass, and stellar population properties.
Due to the poor sky localization of LISA GW detections (typically of order a degree or larger), the number of galaxies within the error volume will be very large \citep{Lops23}. Simulations show that it is very challenging to uniquely identify the host galaxies of GW sources. This is because the preferred host galaxies are dwarfs with merger signatures, but a large fraction of the dwarf galaxy population will exhibit very similar signatures \citep{Villalba}. 

However, we have found that QPE sources (as well as TDEs, Wevers et al. submitted) are strongly overrepresented in galaxies hosting recently faded AGNs and large EELRs with peculiar kinematical properties. Galaxies with similar properties are very rare among the general galaxy population, and hence if the association of QPEs with stellar-mass EMRIs can be confirmed, these peculiar properties can be used to improve the fidelity of EMRI host galaxy localization, which will help to maximize the scientific return of GW astronomy. 

Because the X-ray eruptions have a low duty cycle \citep{Guolo2023} and potentially short lifetimes ($<$1000 years, e.g. \citealt{Linial23}), it is likely that QPEs will be detected in only a small subset of GW detections (see also e.g. \citealt{Franchini23}, who show that only $\sim$2\% of BH EMRIs would produce detectable QPEs). Making associations through their host galaxies can therefore be a valuable complementary method to find EM counterparts.

\section{Conclusions}
\label{sec:summary}
We report on the analysis of MUSE integral field spectroscopy of the host galaxies of five quasi-periodic X-ray eruption (QPE) sources. The sensitivity, large field of view and small spaxel size of MUSE provide an unprecedented, spatially resolved view of the QPE host galaxies. We measure the stellar and emission line kinematics, fluxes and flux ratios to characterize their properties. 

We find that three out of five galaxies have extended emission line regions that have low velocities which are decoupled from the stellar motions. The emission line ratios require a hard non-stellar continuum as the source of the ionizing photons, but the very narrow line velocity dispersions and low velocities are inconsistent with these EELRs being driven by a currently active nucleus. Furthermore, no broad emission lines are observed, which together with the lack of significant obscuration could suggest that no classical BLR region is present\footnote{We note that our observations do not have sufficient spatial resolution to robustly rule out the presence of faint broad emission lines.}. 

The present-day nuclear luminosity estimated from FIR observations is insufficient to satisfy the EELR energy budget, implying that the putative AGNs in these nuclei have decreased their luminosity output significantly in the recent past (9\,000--30\,000 yrs, limited by the extent of the EELR). This could suggest a connection between either the intrinsic rate or the detection rate of QPEs and the specific phase in the AGN duty cycle. The overrepresentation of EELRs in the QPE host galaxies is roughly an order of magnitude compared to post-starburst galaxies, and in combination with the peculiar kinematics, potentially even higher compared to the general galaxy population. 

Several properties of the QPE host galaxies are similar to those of TDE host galaxies. Both classes have a preference for QBS/PSB galaxies \citep{French16, Wevers22}, a strong overrepresentation of EELRs with recently faded nuclei, and peculiar ionized gas kinematic properties (studied in detail for TDE hosts in a companion paper, Wevers et al. submitted). The rarity of the preferred type of host galaxies is highly unlikely to be due to chance, and is robust against associations based on the unique properties of individual QPE sources. This strongly suggests a connection between these two classes of events, consistent with theoretical arguments and hinted at in the observational properties of individual sources \citep{Miniutti23, Arcodia24}. 

The proposed link between QPEs and TDEs disfavors AGN accretion disk models as the origin of QPEs, as i) the AGNs are currently inactive, and ii) this would be the result of two independent selection biases (a detection bias for QPEs, and a detection and/or classification bias for TDEs). The instabilities proposed as the origin for QPEs are typically not "standard" thermal instabilities. Most require some fine-tuning (e.g. magnetic fields inducing very small unstable regions, or independently precessing rings in the case of disc tearing) that are not expected in long-lived large AGN-like discs. 

More detailed studies, including deep narrow-band imaging and/or higher spatial resolution IFU observations, of both QPE and TDE host galaxies can be used to further test the similarity of the host galaxy preferences of these transients. Larger samples can be used to further characterize the properties of these host galaxies and predictions for EMRI GW sources. If these galaxies can be typed by machine learning algorithms, this may facilitate the localization of low frequency GW detections with future observatories.

\begin{acknowledgments}
We are grateful to D. Kakkad, W. Lu, and S. van Velzen for discussions and suggestions, to M. Giustini for providing a recent quiescent luminosity estimate of RXJ1301, and to J. Depasquale for creating the composite color images in Figure \ref{fig:rgb}.
KDF acknowledges support from NSF grant AAG 22-06164. GM was supported by grant PID2020-115325GB-C31 funded by MICIN/AEI/10.13039/501100011033. AIZ
acknowledges support from NASA ADAP grant 80NSSC21K0988. IL acknowledges support from a Rothschild Fellowship and The Gruber Foundation. 
Based on observations collected at the European Southern Observatory under ESO programmes 109.238W.007, 111.24UJ.005 and 113.26F6.001.
This research was supported in part by grant NSF PHY-2309135 to the Kavli Institute for Theoretical Physics (KITP). We thank the organizers of the KITP program: Towards a Physical Understanding of Tidal Disruption Events, where part of this work was performed. 
\end{acknowledgments}
\vspace{5mm}
\facilities{ESO/VLT}

\software{astropy \citep{2013A&A...558A..33A,2018AJ....156..123A}}

\appendix
\setcounter{table}{0}
\renewcommand{\thetable}{A\arabic{table}}
\setcounter{figure}{0}
\renewcommand{\thefigure}{A\arabic{figure}}

\section*{Supplementary material}

\begin{figure*}
    \centering
    \includegraphics[width=0.47\textwidth]{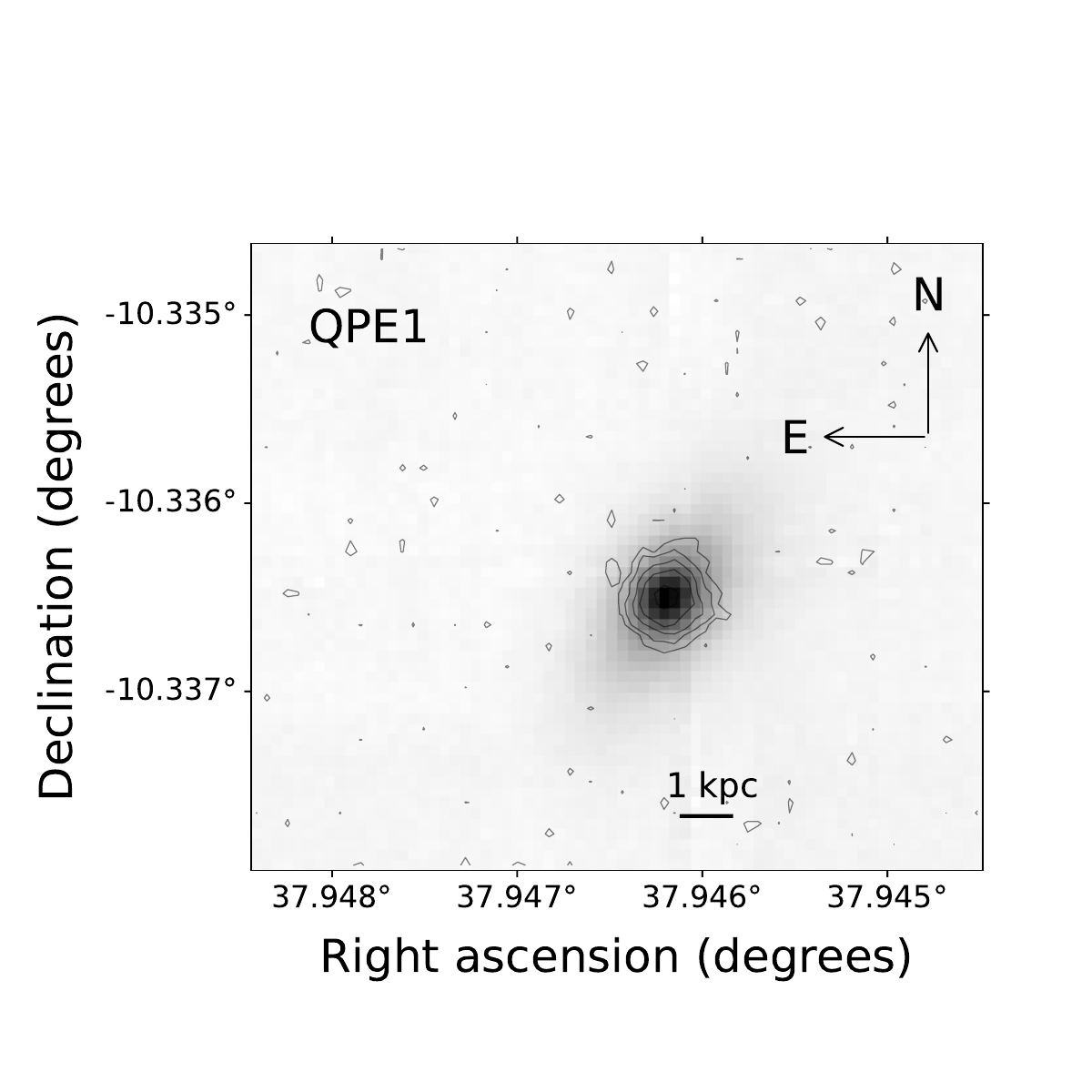}
    \includegraphics[width=0.47\textwidth]{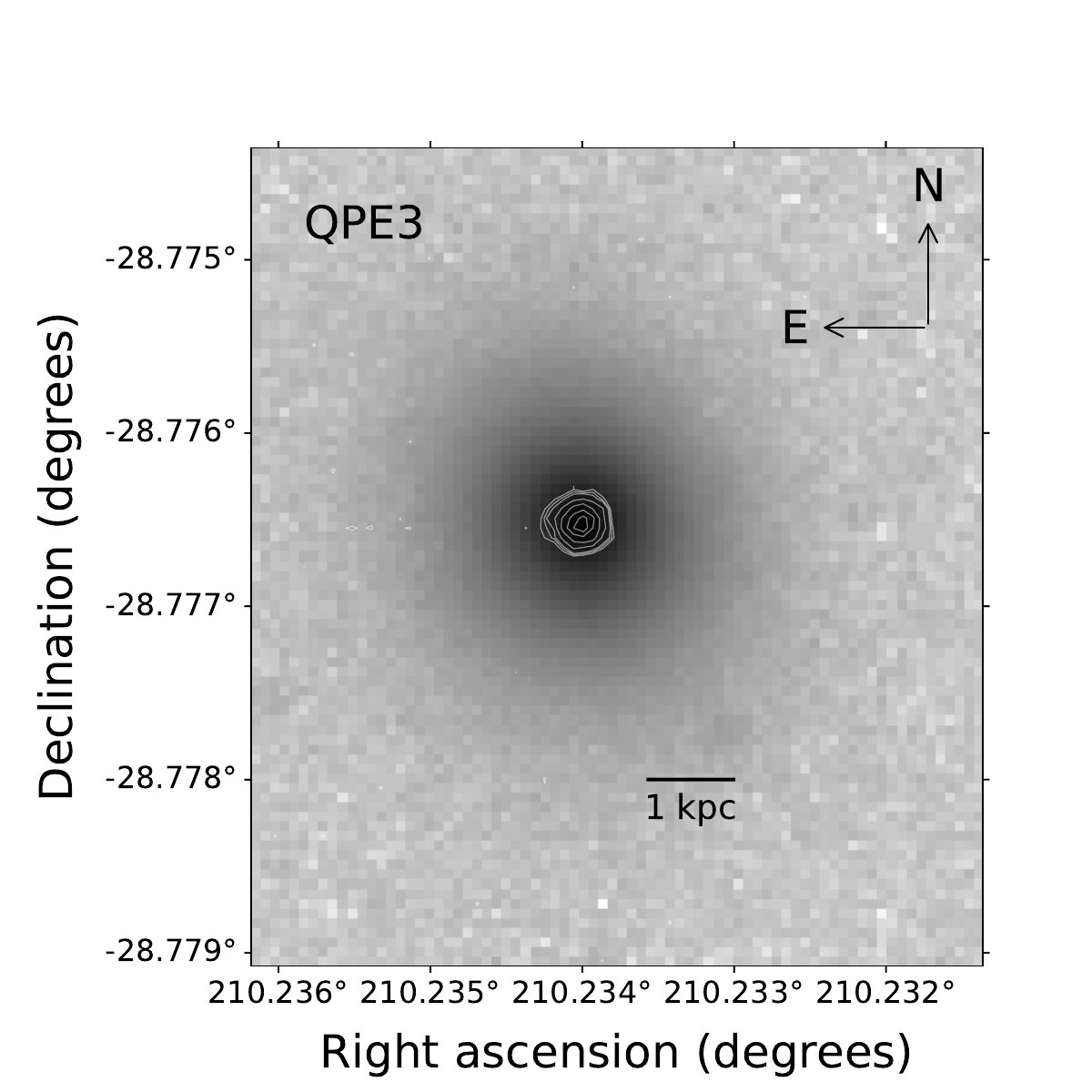}
    \caption{Continuum grayscale image and H$\alpha$ contours overlaid for the source eRO-QPE1 (left) and eRO-QPE3 (right), which do not show an EELR. The faintest contours represent H$\alpha$ flux levels of 35$\times$ 10$^{-20}$ erg cm$^{-2}$ s$^{-1}$ \AA\ and 15$\times$ 10$^{-20}$ erg cm$^{-2}$ s$^{-1}$ \AA\ for eRO-QPE1 and eRO-QPE3, respectively.}
    \label{fig:overlay_qpe1}
\end{figure*}

\begin{figure}
    \centering
    \includegraphics[width=0.47\textwidth]{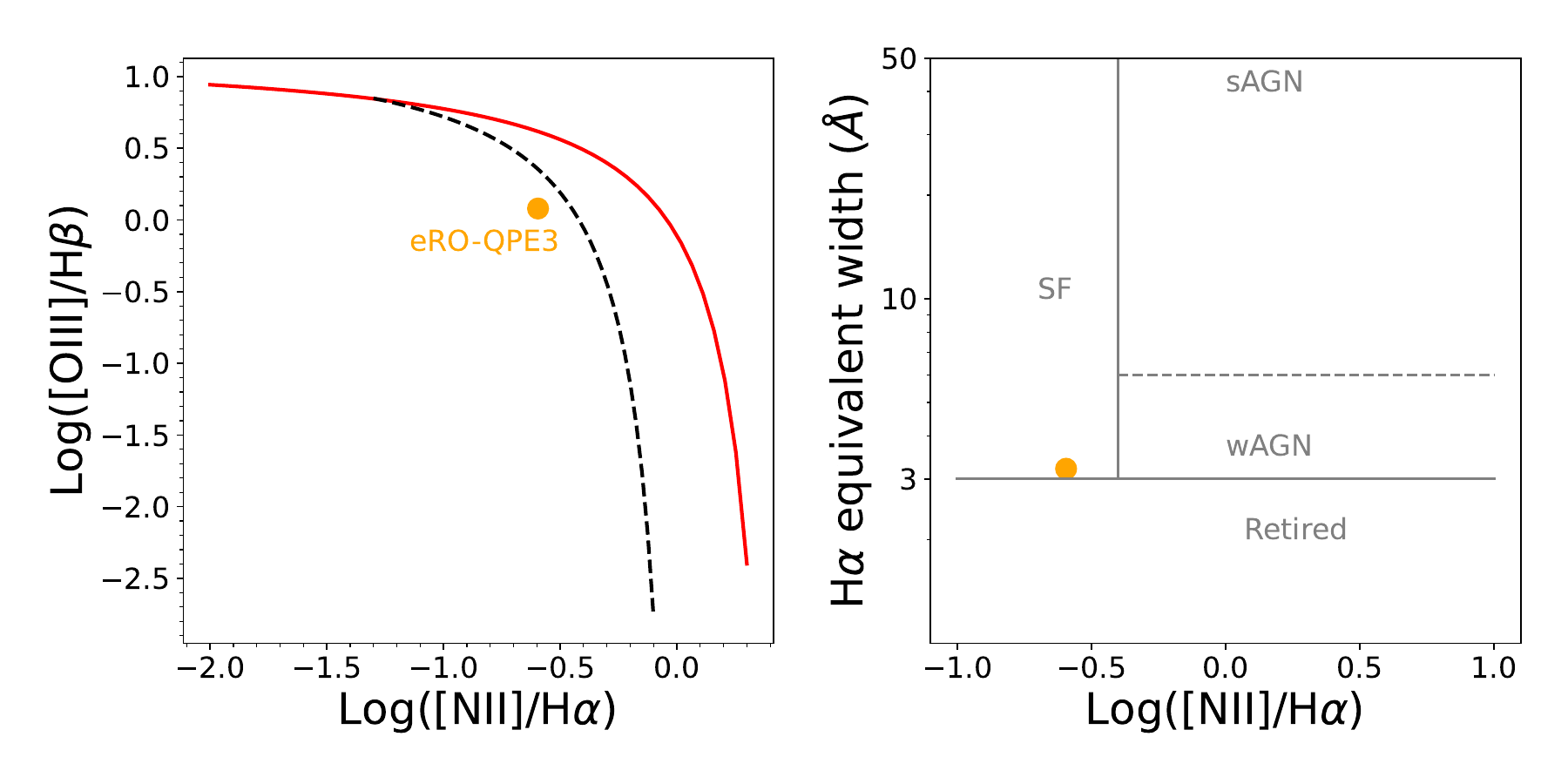}
    \caption{BPT and WHAN diagram for the source eRO-QPE3.}
    \label{fig:qpe3}
\end{figure}

\begin{figure*}
    \centering
    \includegraphics[width=0.32\textwidth]{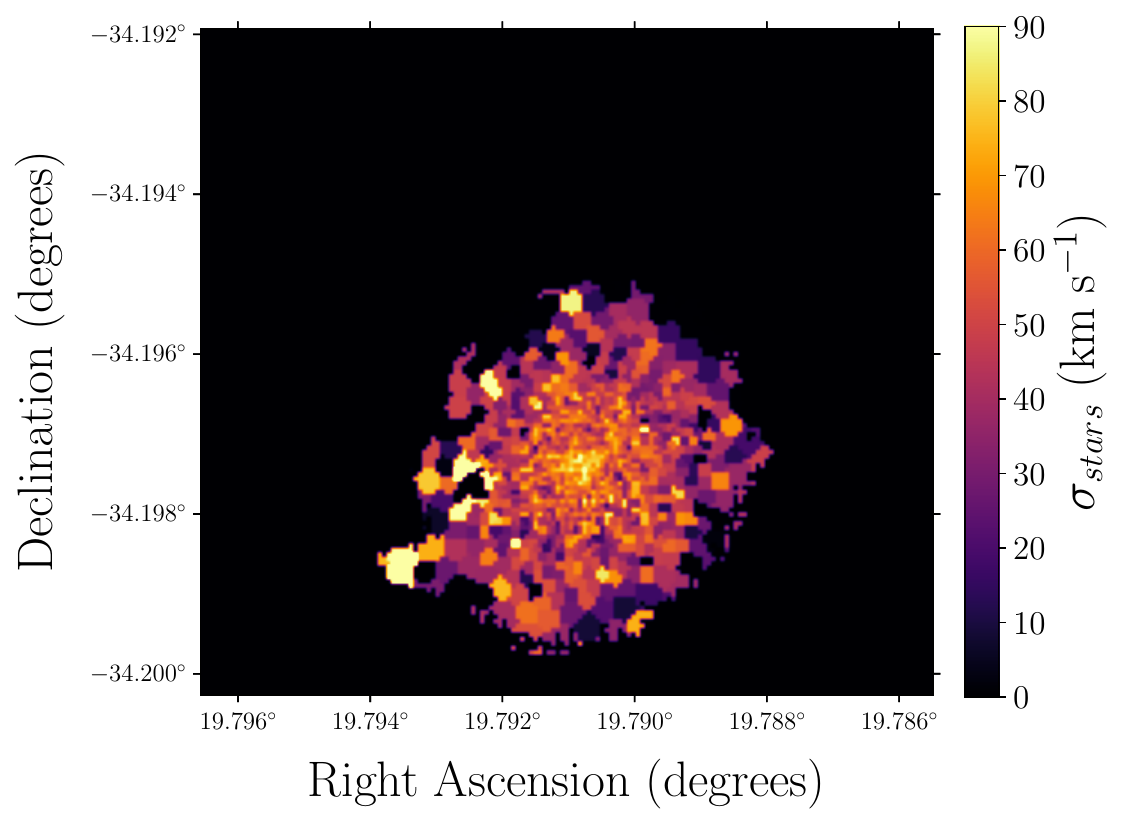}
    \includegraphics[width=0.32\textwidth]{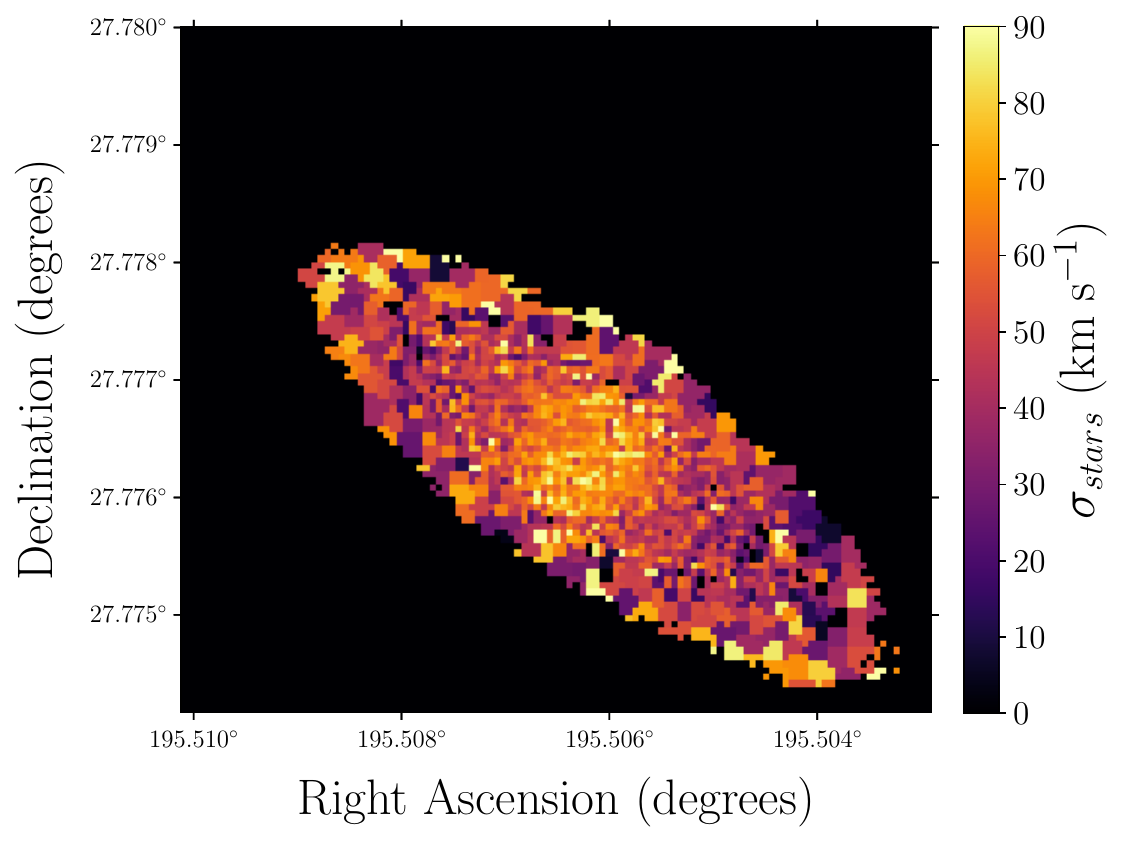}
    \includegraphics[width=0.32\textwidth]{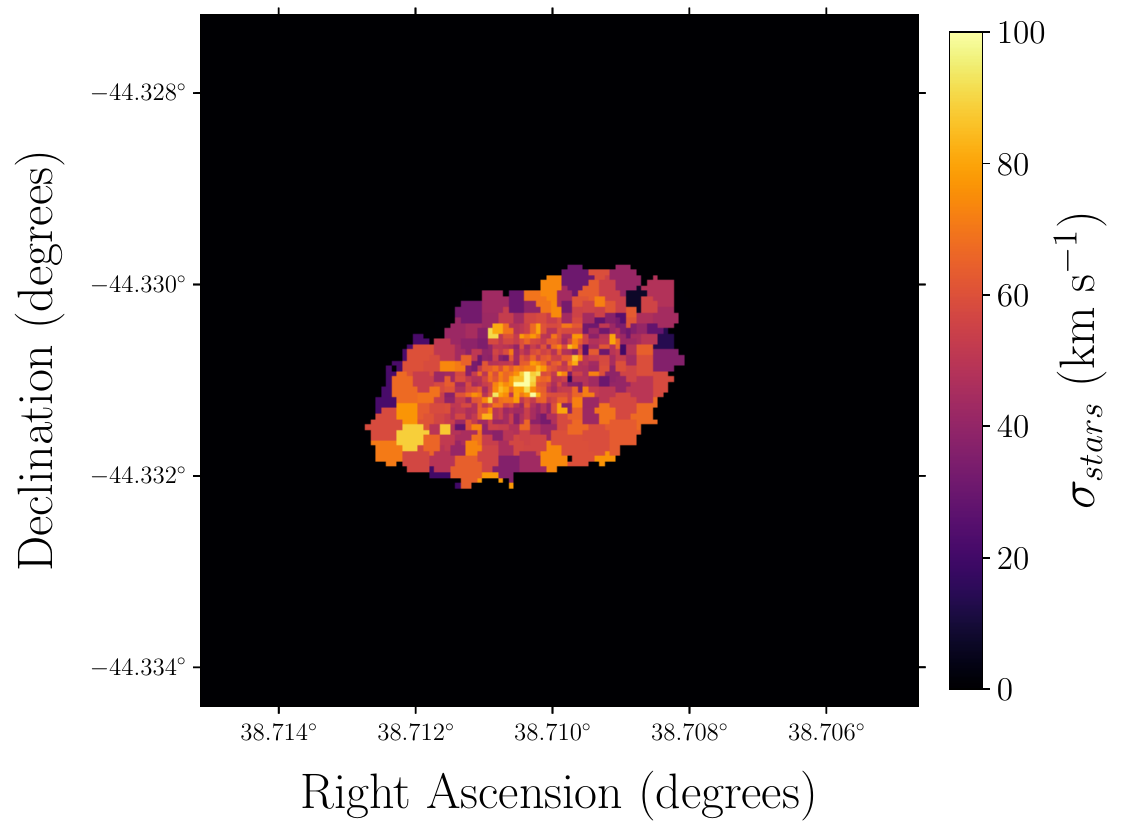}
    \caption{Stellar velocity dispersion maps of GSN069 (left), RXJ1301 (middle) and eRO-QPE2 (right).}
    \label{fig:veldisp}
\end{figure*}

\begin{figure*}
    \centering
    \includegraphics[width=\textwidth]{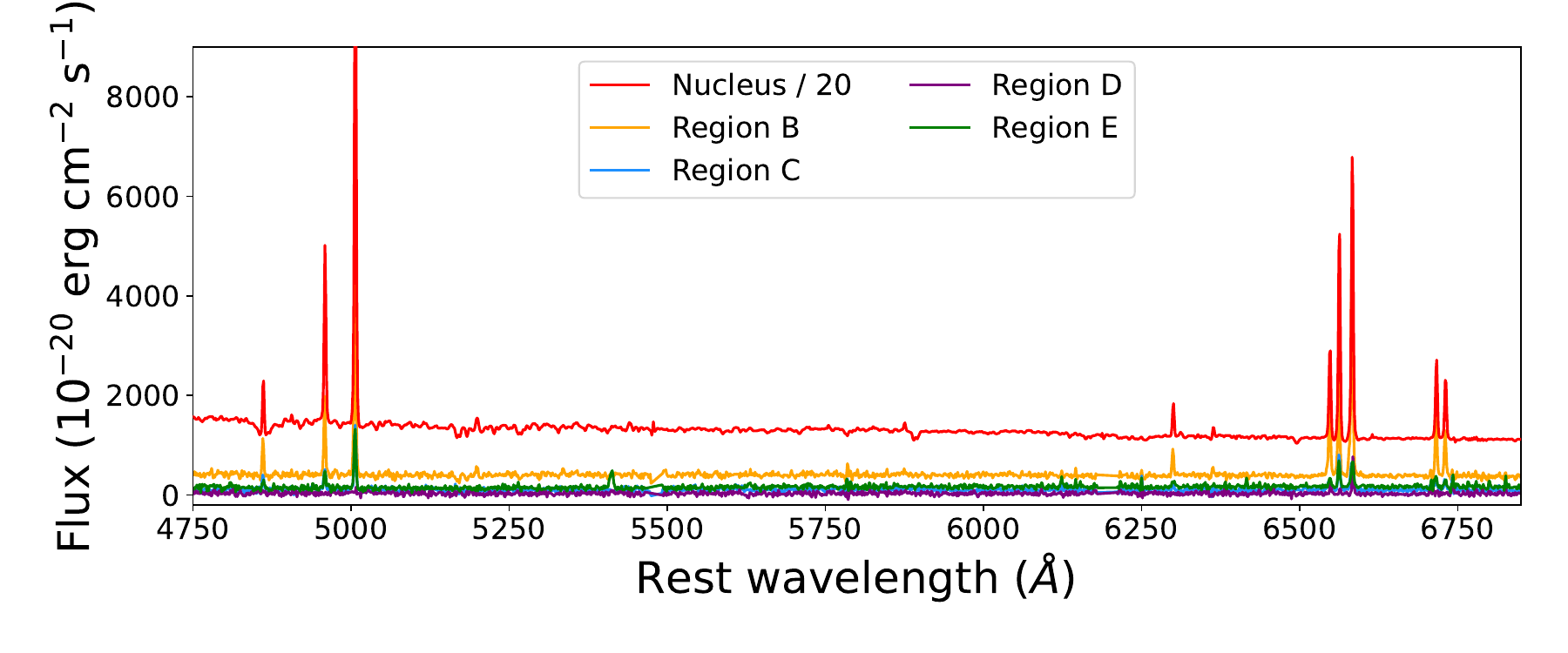}
    \caption{Aperture spectra of GSN069, for the regions indicated in Figure \ref{fig:rgb}, with the same color coding.}
    \label{fig:spectra_gsn069}
\end{figure*}

\begin{figure*}
    \centering
    \includegraphics[width=\textwidth]{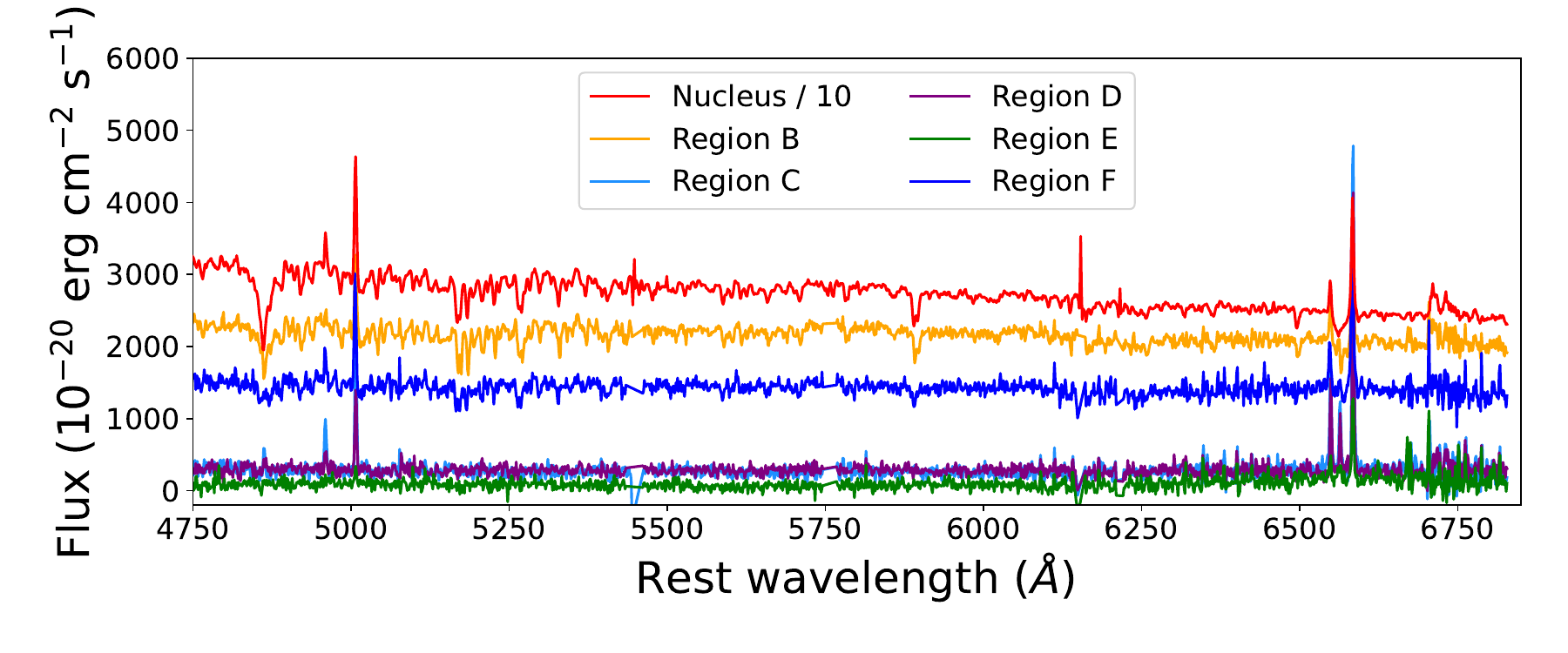}
    \caption{Aperture spectra of RXJ1301, for the regions indicated in Figure \ref{fig:rgb}, with the same color coding.}
    \label{fig:spectra_rxj1301}
\end{figure*}

\begin{figure*}
    \centering
    \includegraphics[width=\textwidth]{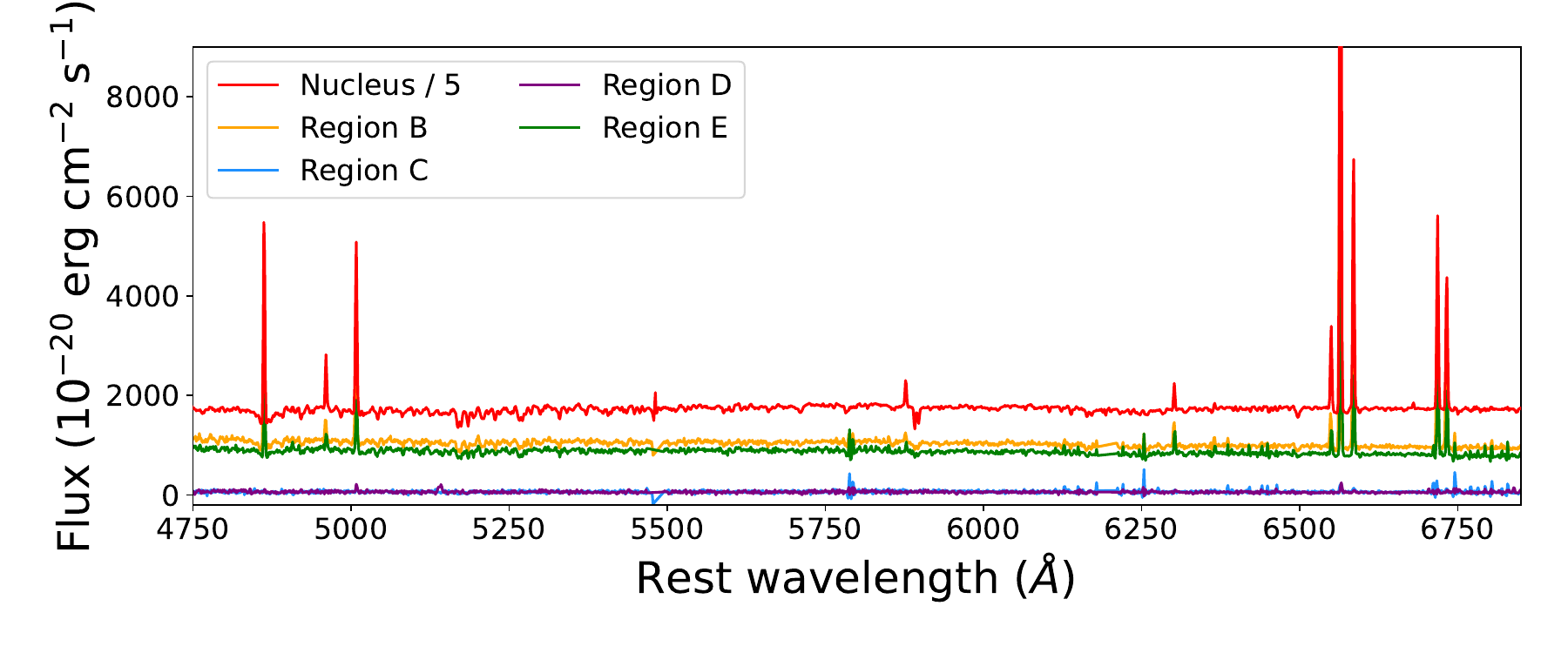}
    \caption{Aperture spectra of eRO-QPE2, for the regions indicated in Figure \ref{fig:rgb}, with the same color coding.}
    \label{fig:spectra_qpe2}
\end{figure*}

\begin{figure*}
    \centering
    \includegraphics[width=0.47\textwidth]{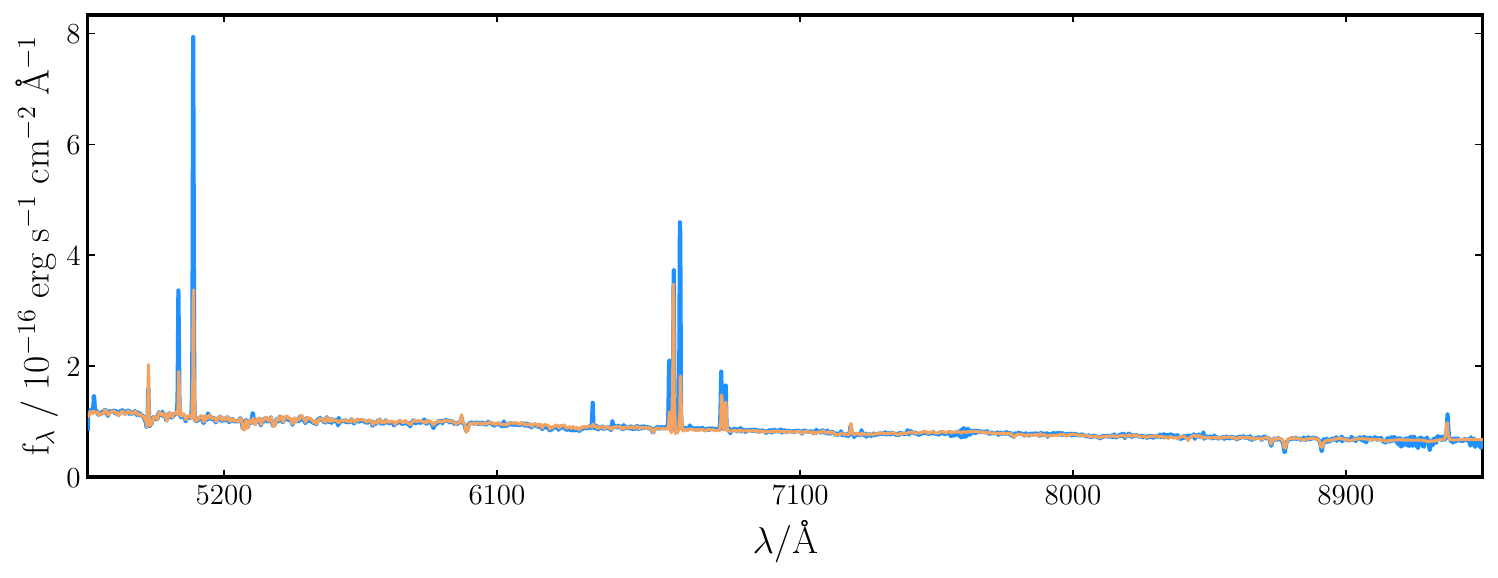}
    \includegraphics[width=0.47\textwidth]{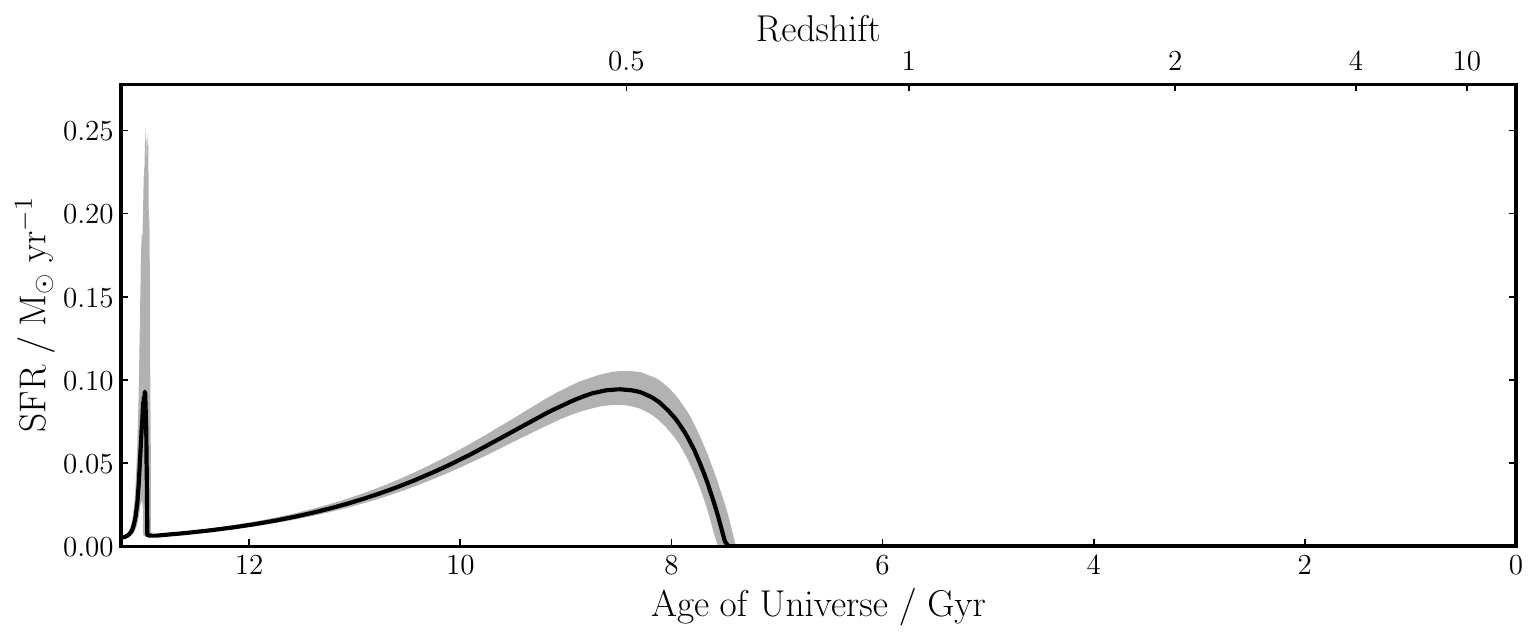}
    \includegraphics[width=0.47\textwidth]{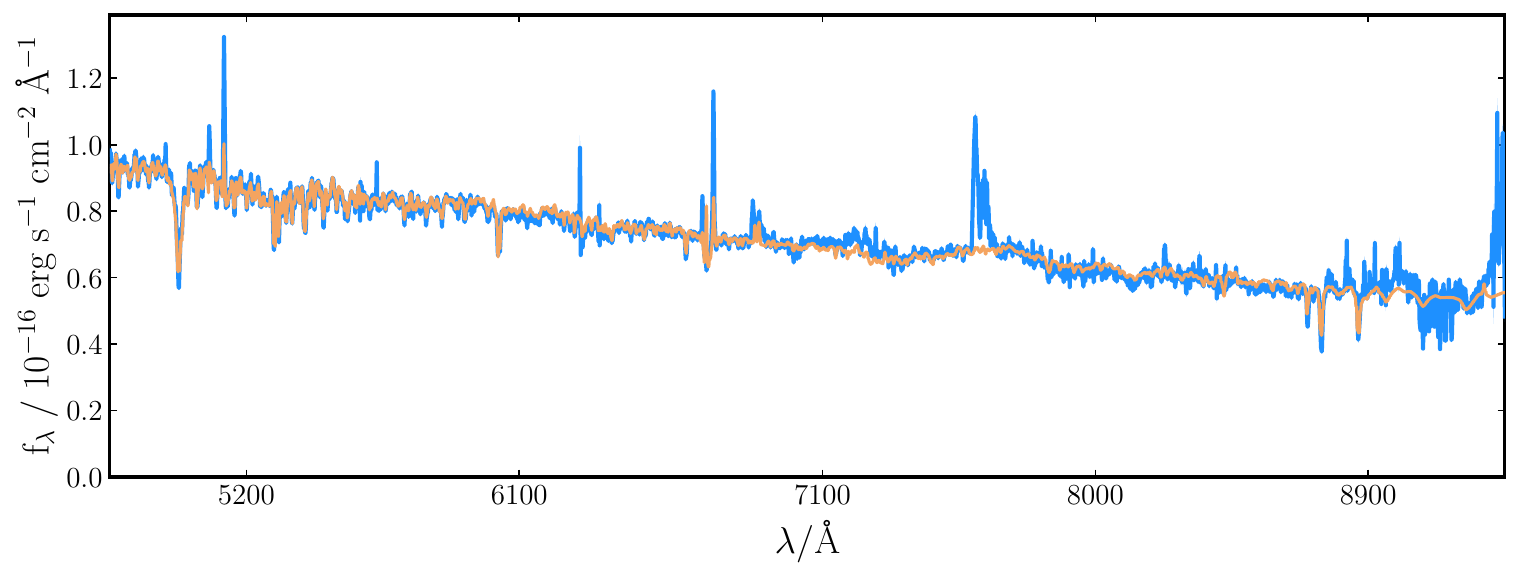}
    \includegraphics[width=0.47\textwidth]{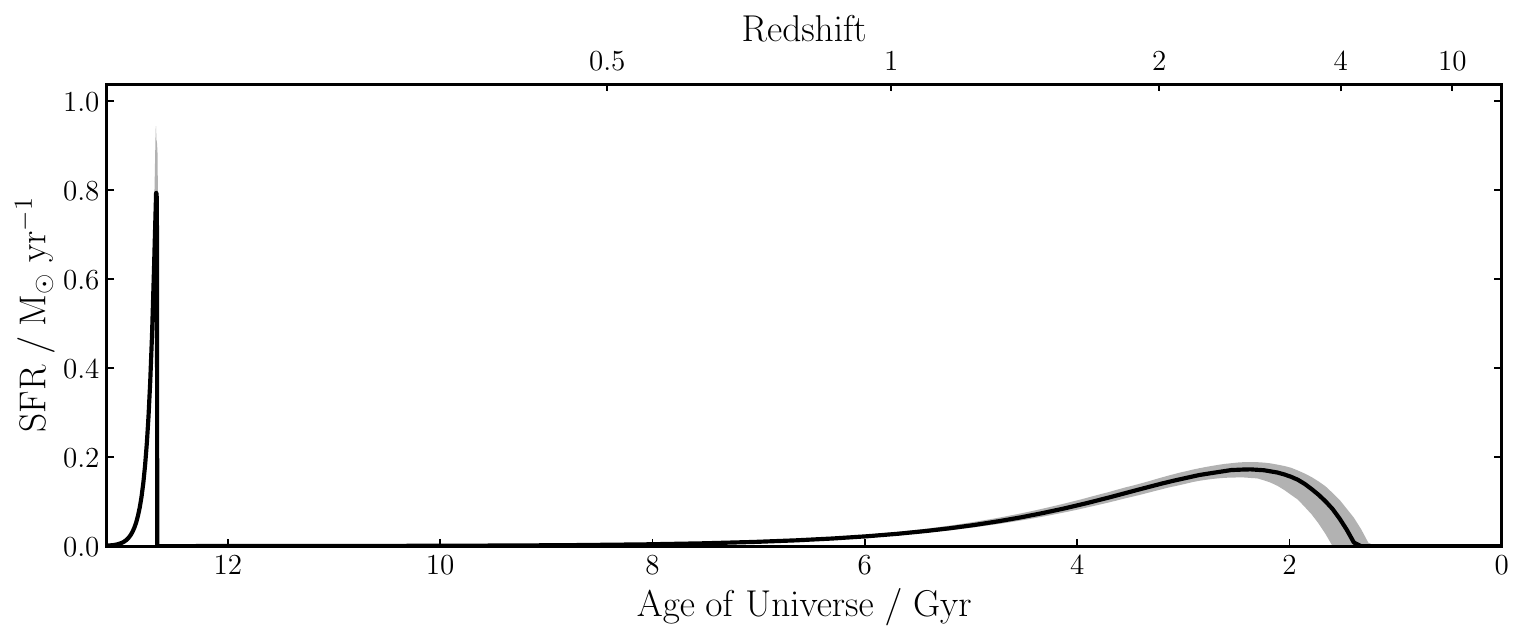}
    \includegraphics[width=0.47\textwidth]{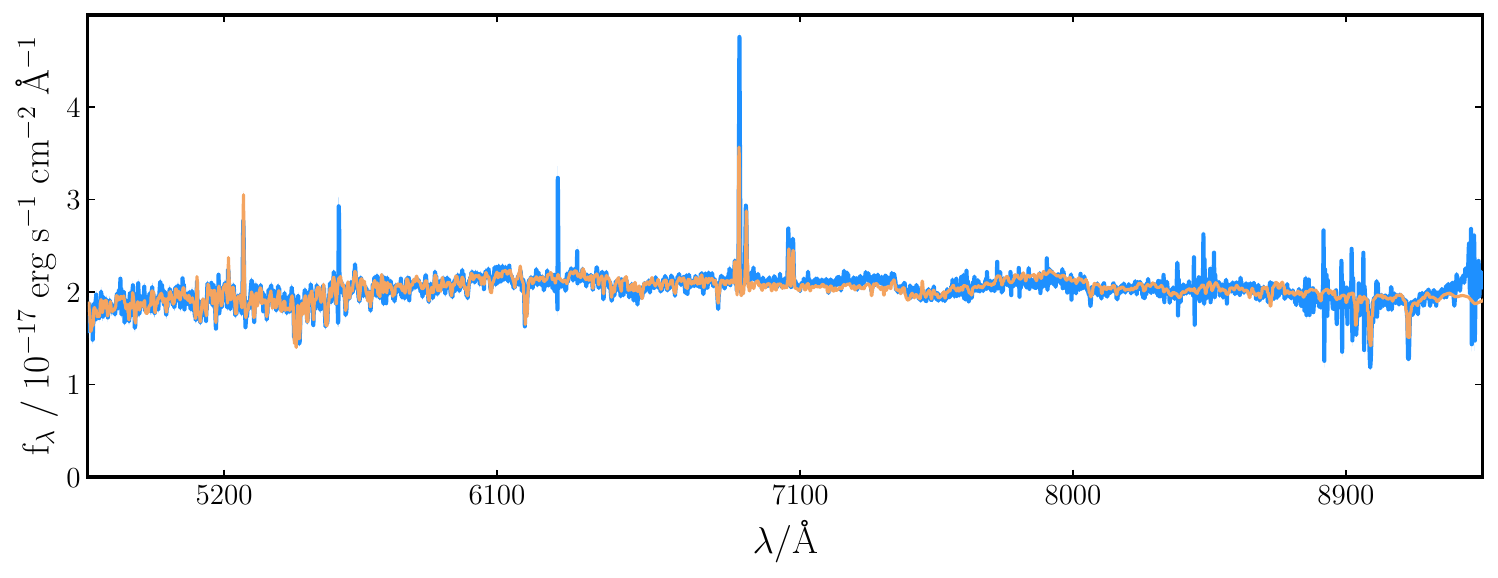}
    \includegraphics[width=0.47\textwidth]{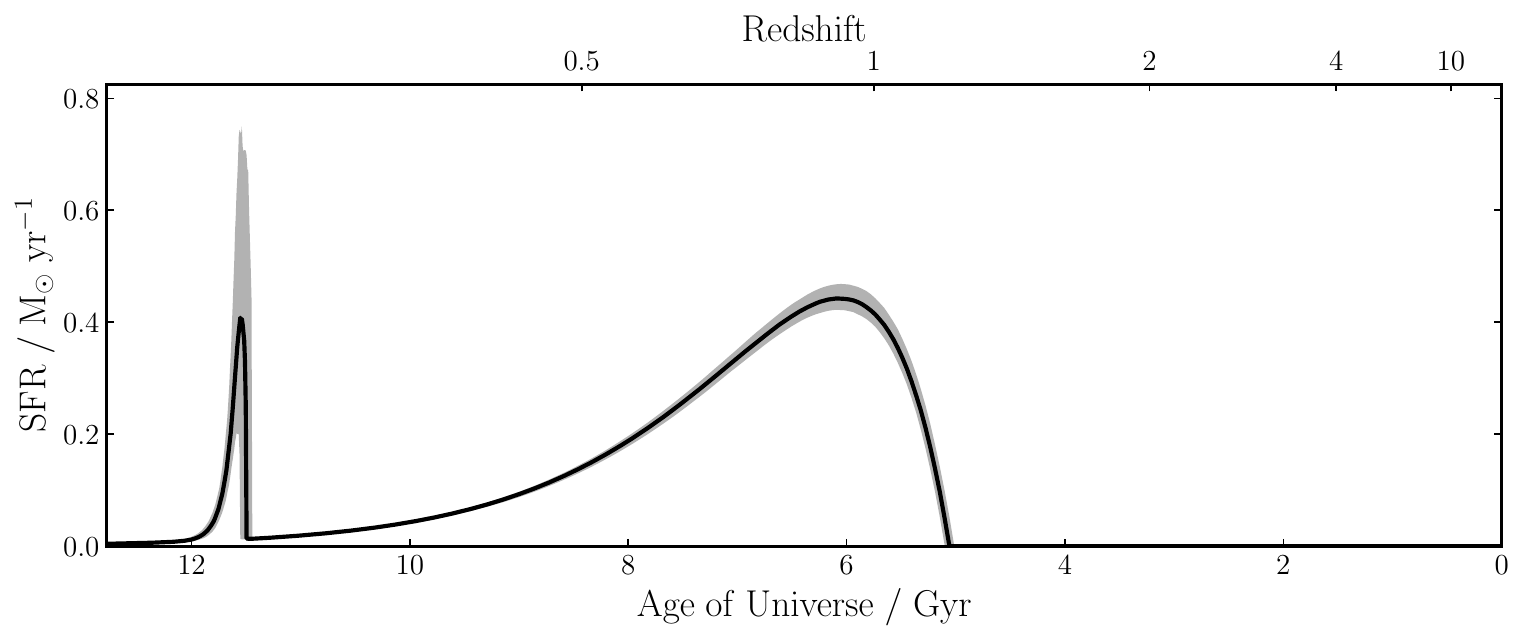}
    \includegraphics[width=0.47\textwidth]{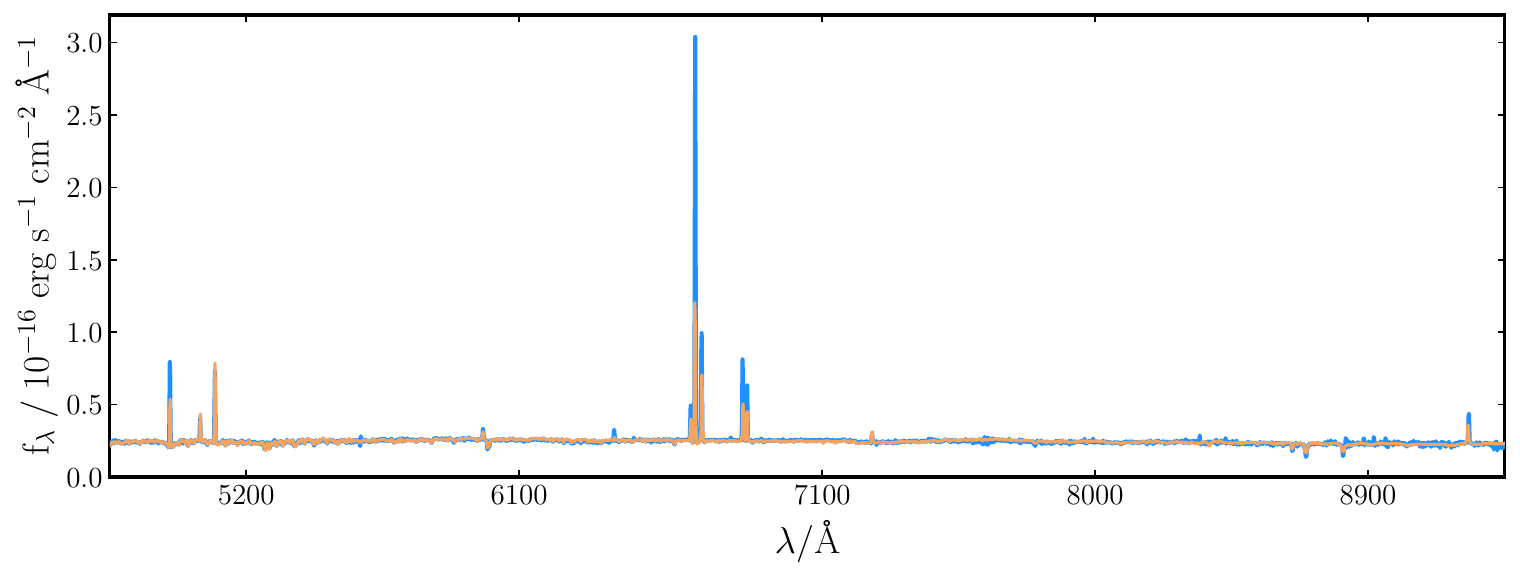}
    \includegraphics[width=0.47\textwidth]{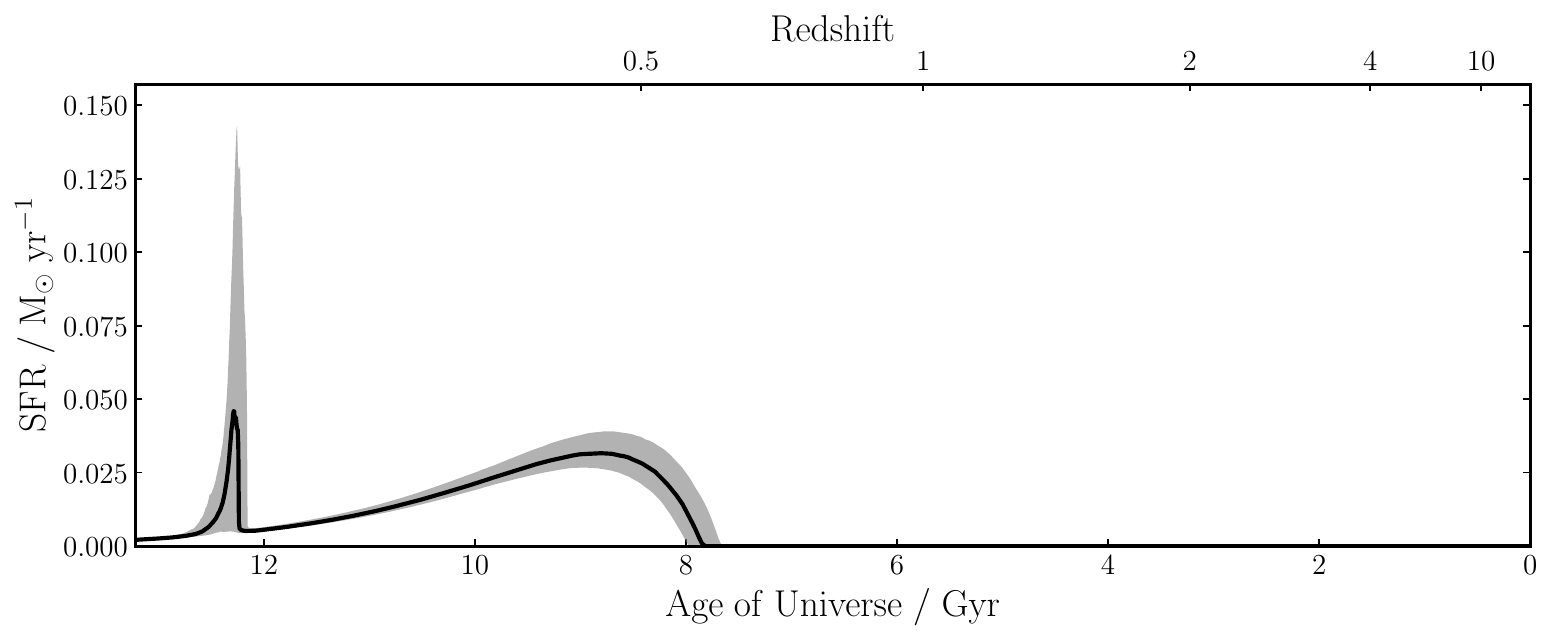}
    \includegraphics[width=0.47\textwidth]{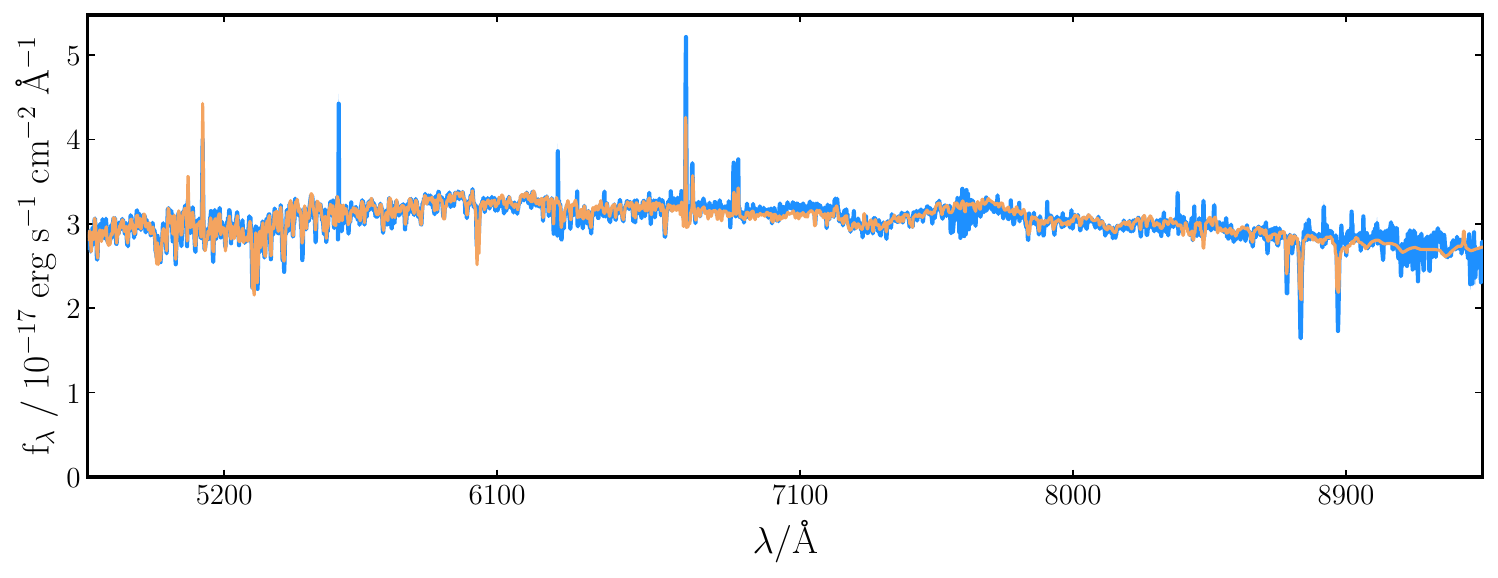}
    \includegraphics[width=0.47\textwidth]{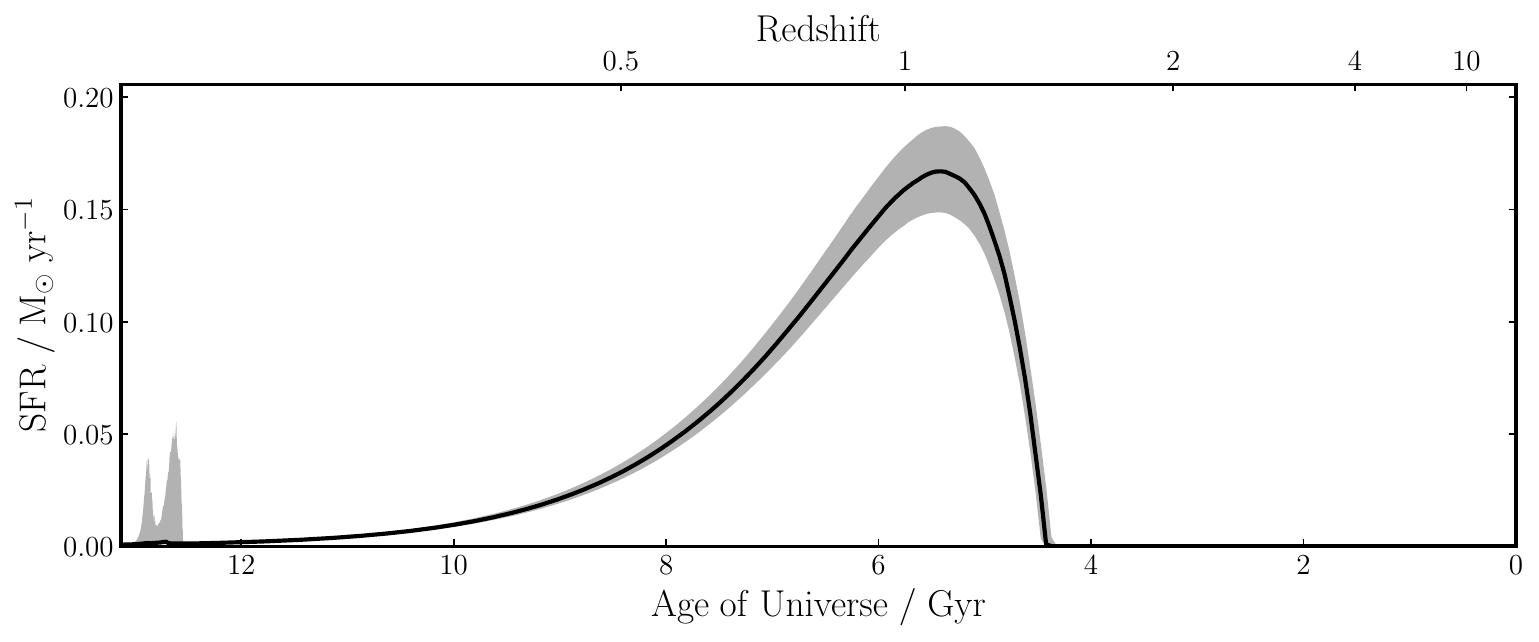}
    \caption{Bagpipes fits and inferred star formation history, assuming a double starburst as detailed in the text. From top to bottom: GSN069, RXJ1301, eRO-QPE1, eRO-QPE2, and eRO-QPE3.}
    \label{fig:bagpipes}
\end{figure*}

\begin{table}
    \caption{Overview of the QPE host galaxies analyzed in this work. Spatial resolution refers to the native 0.2 arcsec pixel scale of the MUSE WFM-NOAO observations. The PSF FWHM is estimated using point sources in the continuum cube when available, or otherwise taken as the average seeing conditions (uncorrected for airmass) provided by the observatory (marked with an asterisk) as tabulated in Table \ref{tab:muse}. $\sigma_{\rm star}$ provides the central velocity dispersion, measured using an aperture with the size of the PSF FWHM.}
    \centering
    \begin{tabular}{c|ccccccc}
    Source & $z$ & Spatial resolution & PSF FWHM & PSF FWHM &  E(B--V)$_{\rm gal}$ & $\sigma_{\rm star}$ & Reference\\
     & & (pc / pix) & (arcsec) & (pc) & (mag) & km s$^{-1}$ \\\hline
        GSN069  & 0.0182& 74 & 1.0 & 370 &  0.0232 & 68$\pm$12 & \citet{Saxton11}\\ 
        RXJ1301 & 0.0237& 96 &0.94$^*$  & 450 &  0.0077 & 68$\pm$5 & \citet{Dewangan2000}\\
        eRO-QPE1 & 0.0505& 199 &0.95$^*$ & 940 & 0.0212 & 66$\pm$4 & \citet{Arcodia21}\\
        eRO-QPE2 & 0.0175& 71 & 0.7 & 250 &  0.0153 & 38$\pm$6 &  \citet{Arcodia21}\\
        eRO-QPE3 & 0.024 & 97 & 0.7 & 340 & 0.0505 & 38$\pm$5 & \citet{Arcodia24}\\\hline
    \end{tabular}
    \label{tab:observations}
\end{table}

\begin{table*}
    \caption{Observing log of the VLT/MUSE data. All data was taken in Wide Field no-AO Mode with the nominal filter.}
    \centering
    \begin{tabular}{c|cccc}
    Source & Date & Exposure time & Airmass & DIMM seeing\\
     & (MJD) & (Seconds) & & (arcsec) \\\hline
       GSN069  & 59787.237738 & 1497&1.64 & 0.53 \\
       & 59787.256443 & 1497&1.46 & 0.73 \\ 
        RXJ1301 & 59787.966968 & 1497&1.92 & 1.10  \\
        & 59787.985819	& 1497&2.11 & 0.78 \\
        eRO-QPE1 & 59834.144255 & 1497& 2.25 & 1.11 \\
        & 59834.162944 & 1497&1.86 & 0.75 \\
        & 59834.263008 & 1497&1.11 & 1.03 \\
        & 59834.281717& 1497&1.07 & 1.04 \\
        & 59834.303142& 1497&1.04 & 0.74 \\
        &59834.321856 & 1497&1.03 & 0.87  \\
        & 59834.343499	& 1497&1.04 & 1.04 \\
        & 59834.362185 & 1497&1.06 & 0.98 \\
        eRO-QPE2 & 59787.369579& 1497& 1.18 & 0.63  \\
        & 59787.388267 & 1497& 1.13 & 0.51 \\
        & 59814.263422& 1497& 1.30 &2.19 \\
        & 59814.282118& 1497& 1.23 & 0.97 \\
        eRO-QPE3 & 60416.343266 & 1497 & 1.352 & 4.84 \\	
        & 60416.361829	& 1498 & 1.498 & 4.84 \\
        & 60432.108992	& 1492 & 1.060	& 0.33 \\
        & 60432.127534	& 1493 & 1.029	& 0.44 \\
        & 60435.208337	& 1480 & 1.041 &  1.21 \\
        & 60435.226639	& 1480 & 1.078	& 1.16 \\
        & 60438.139092	& 1487 & 1.005	& 0.81 \\
        & 60438.157507	& 1488 & 1.003	& 0.64 \\
        & 60439.167192	& 1486 & 1.008	& 0.78 \\
        & 60439.185699	& 1486 & 1.025	& 0.85\\ \hline
    \end{tabular}
    \label{tab:muse}
\end{table*}
\bibliography{sample631}

\begin{thebibliography}{}
\expandafter\ifx\csname natexlab\endcsname\relax\def\natexlab#1{#1}\fi
\providecommand{\url}[1]{\href{#1}{#1}}
\providecommand{\dodoi}[1]{doi:~\href{http://doi.org/#1}{\nolinkurl{#1}}}
\providecommand{\doeprint}[1]{\href{http://ascl.net/#1}{\nolinkurl{http://ascl.net/#1}}}
\providecommand{\doarXiv}[1]{\href{https://arxiv.org/abs/#1}{\nolinkurl{https://arxiv.org/abs/#1}}}

\bibitem[{{Arcodia} {et~al.}(2021){Arcodia}, {Merloni}, {Nandra}, {Buchner},
  {Salvato}, {Pasham}, {Remillard}, {Comparat}, {Lamer}, {Ponti}, {Malyali},
  {Wolf}, {Arzoumanian}, {Bogensberger}, {Buckley}, {Gendreau}, {Gromadzki},
  {Kara}, {Krumpe}, {Markwardt}, {Ramos-Ceja}, {Rau}, {Schramm}, \&
  {Schwope}}]{Arcodia21}
{Arcodia}, R., {Merloni}, A., {Nandra}, K., {et~al.} 2021, \nat, 592, 704,
  \dodoi{10.1038/s41586-021-03394-6}

\bibitem[{{Arcodia} {et~al.}(2024){Arcodia}, {Liu}, {Merloni}, {Malyali},
  {Rau}, {Chakraborty}, {Goodwin}, {Buckley}, {Brink}, {Gromadzki},
  {Arzoumanian}, {Buchner}, {Kara}, {Nandra}, {Ponti}, {Salvato}, {Anderson},
  {Baldini}, {Grotova}, {Krumpe}, {Maitra}, {Miller-Jones}, \&
  {Ramos-Ceja}}]{Arcodia24}
{Arcodia}, R., {Liu}, Z., {Merloni}, A., {et~al.} 2024, \aap, 684, A64,
  \dodoi{10.1051/0004-6361/202348881}

\bibitem[{{Astropy Collaboration} {et~al.}(2013){Astropy Collaboration},
  {Robitaille}, {Tollerud}, {Greenfield}, {Droettboom}, {Bray}, {Aldcroft},
  {Davis}, {Ginsburg}, {Price-Whelan}, {Kerzendorf}, {Conley}, {Crighton},
  {Barbary}, {Muna}, {Ferguson}, {Grollier}, {Parikh}, {Nair}, {Unther},
  {Deil}, {Woillez}, {Conseil}, {Kramer}, {Turner}, {Singer}, {Fox}, {Weaver},
  {Zabalza}, {Edwards}, {Azalee Bostroem}, {Burke}, {Casey}, {Crawford},
  {Dencheva}, {Ely}, {Jenness}, {Labrie}, {Lim}, {Pierfederici}, {Pontzen},
  {Ptak}, {Refsdal}, {Servillat}, \& {Streicher}}]{2013A&A...558A..33A}
{Astropy Collaboration}, {Robitaille}, T.~P., {Tollerud}, E.~J., {et~al.} 2013,
  \aap, 558, A33, \dodoi{10.1051/0004-6361/201322068}

\bibitem[{{Astropy Collaboration} {et~al.}(2018){Astropy Collaboration},
  {Price-Whelan}, {Sip{\H{o}}cz}, {G{\"u}nther}, {Lim}, {Crawford}, {Conseil},
  {Shupe}, {Craig}, {Dencheva}, {Ginsburg}, {VanderPlas}, {Bradley},
  {P{\'e}rez-Su{\'a}rez}, {de Val-Borro}, {Aldcroft}, {Cruz}, {Robitaille},
  {Tollerud}, {Ardelean}, {Babej}, {Bach}, {Bachetti}, {Bakanov}, {Bamford},
  {Barentsen}, {Barmby}, {Baumbach}, {Berry}, {Biscani}, {Boquien}, {Bostroem},
  {Bouma}, {Brammer}, {Bray}, {Breytenbach}, {Buddelmeijer}, {Burke},
  {Calderone}, {Cano Rodr{\'\i}guez}, {Cara}, {Cardoso}, {Cheedella}, {Copin},
  {Corrales}, {Crichton}, {D'Avella}, {Deil}, {Depagne}, {Dietrich}, {Donath},
  {Droettboom}, {Earl}, {Erben}, {Fabbro}, {Ferreira}, {Finethy}, {Fox},
  {Garrison}, {Gibbons}, {Goldstein}, {Gommers}, {Greco}, {Greenfield},
  {Groener}, {Grollier}, {Hagen}, {Hirst}, {Homeier}, {Horton}, {Hosseinzadeh},
  {Hu}, {Hunkeler}, {Ivezi{\'c}}, {Jain}, {Jenness}, {Kanarek}, {Kendrew},
  {Kern}, {Kerzendorf}, {Khvalko}, {King}, {Kirkby}, {Kulkarni}, {Kumar},
  {Lee}, {Lenz}, {Littlefair}, {Ma}, {Macleod}, {Mastropietro}, {McCully},
  {Montagnac}, {Morris}, {Mueller}, {Mumford}, {Muna}, {Murphy}, {Nelson},
  {Nguyen}, {Ninan}, {N{\"o}the}, {Ogaz}, {Oh}, {Parejko}, {Parley}, {Pascual},
  {Patil}, {Patil}, {Plunkett}, {Prochaska}, {Rastogi}, {Reddy Janga},
  {Sabater}, {Sakurikar}, {Seifert}, {Sherbert}, {Sherwood-Taylor}, {Shih},
  {Sick}, {Silbiger}, {Singanamalla}, {Singer}, {Sladen}, {Sooley},
  {Sornarajah}, {Streicher}, {Teuben}, {Thomas}, {Tremblay}, {Turner},
  {Terr{\'o}n}, {van Kerkwijk}, {de la Vega}, {Watkins}, {Weaver}, {Whitmore},
  {Woillez}, {Zabalza}, \& {Astropy Contributors}}]{2018AJ....156..123A}
{Astropy Collaboration}, {Price-Whelan}, A.~M., {Sip{\H{o}}cz}, B.~M., {et~al.}
  2018, \aj, 156, 123, \dodoi{10.3847/1538-3881/aabc4f}

\bibitem[{{Bacon} {et~al.}(2010){Bacon}, {Accardo}, {Adjali}, {Anwand},
  {Bauer}, {Biswas}, {Blaizot}, {Boudon}, {Brau-Nogue}, {Brinchmann},
  {Caillier}, {Capoani}, {Carollo}, {Contini}, {Couderc}, {Daguis{\'e}},
  {Deiries}, {Delabre}, {Dreizler}, {Dubois}, {Dupieux}, {Dupuy}, {Emsellem},
  {Fechner}, {Fleischmann}, {Fran{\c{c}}ois}, {Gallou}, {Gharsa}, {Glindemann},
  {Gojak}, {Guiderdoni}, {Hansali}, {Hahn}, {Jarno}, {Kelz}, {Koehler},
  {Kosmalski}, {Laurent}, {Le Floch}, {Lilly}, {Lizon}, {Loupias}, {Manescau},
  {Monstein}, {Nicklas}, {Olaya}, {Pares}, {Pasquini}, {P{\'e}contal-Rousset},
  {Pell{\'o}}, {Petit}, {Popow}, {Reiss}, {Remillieux}, {Renault}, {Roth},
  {Rupprecht}, {Serre}, {Schaye}, {Soucail}, {Steinmetz}, {Streicher}, {Stuik},
  {Valentin}, {Vernet}, {Weilbacher}, {Wisotzki}, \& {Yerle}}]{Bacon_2010}
{Bacon}, R., {Accardo}, M., {Adjali}, L., {et~al.} 2010, in Society of
  Photo-Optical Instrumentation Engineers (SPIE) Conference Series, Vol. 7735,
  Ground-based and Airborne Instrumentation for Astronomy III, ed. I.~S.
  {McLean}, S.~K. {Ramsay}, \& H.~{Takami}, 773508, \dodoi{10.1117/12.856027}

\bibitem[{{Baron} {et~al.}(2022){Baron}, {Netzer}, {Lutz}, {Prochaska}, \&
  {Davies}}]{Baron22}
{Baron}, D., {Netzer}, H., {Lutz}, D., {Prochaska}, J.~X., \& {Davies}, R.~I.
  2022, \mnras, 509, 4457, \dodoi{10.1093/mnras/stab3232}

\bibitem[{{Cappellari}(2023)}]{Cappellari23}
{Cappellari}, M. 2023, \mnras, 526, 3273, \dodoi{10.1093/mnras/stad2597}

\bibitem[{{Carnall} {et~al.}(2018){Carnall}, {McLure}, {Dunlop}, \&
  {Dav{\'e}}}]{Carnall2018}
{Carnall}, A.~C., {McLure}, R.~J., {Dunlop}, J.~S., \& {Dav{\'e}}, R. 2018,
  \mnras, 480, 4379, \dodoi{10.1093/mnras/sty2169}

\bibitem[{{Carnall} {et~al.}(2019){Carnall}, {McLure}, {Dunlop}, {Cullen},
  {McLeod}, {Wild}, {Johnson}, {Appleby}, {Dav{\'e}}, {Amorin}, {Bolzonella},
  {Castellano}, {Cimatti}, {Cucciati}, {Gargiulo}, {Garilli}, {Marchi},
  {Pentericci}, {Pozzetti}, {Schreiber}, {Talia}, \& {Zamorani}}]{Carnall2019}
{Carnall}, A.~C., {McLure}, R.~J., {Dunlop}, J.~S., {et~al.} 2019, \mnras, 490,
  417, \dodoi{10.1093/mnras/stz2544}

\bibitem[{{Chakraborty} {et~al.}(2021){Chakraborty}, {Kara}, {Masterson},
  {Giustini}, {Miniutti}, \& {Saxton}}]{Chakraborty21}
{Chakraborty}, J., {Kara}, E., {Masterson}, M., {et~al.} 2021, \apjl, 921, L40,
  \dodoi{10.3847/2041-8213/ac313b}

\bibitem[{{Chan} {et~al.}(2019){Chan}, {Piran}, {Krolik}, \&
  {Saban}}]{Chan2019}
{Chan}, C.-H., {Piran}, T., {Krolik}, J.~H., \& {Saban}, D. 2019, \apj, 881,
  113, \dodoi{10.3847/1538-4357/ab2b40}

\bibitem[{{Dewangan} {et~al.}(2000){Dewangan}, {Singh}, {Mayya}, \&
  {Anupama}}]{Dewangan2000}
{Dewangan}, G.~C., {Singh}, K.~P., {Mayya}, Y.~D., \& {Anupama}, G.~C. 2000,
  \mnras, 318, 309, \dodoi{10.1046/j.1365-8711.2000.03755.x}

\bibitem[{{Franchini} {et~al.}(2023){Franchini}, {Bonetti}, {Lupi}, {Miniutti},
  {Bortolas}, {Giustini}, {Dotti}, {Sesana}, {Arcodia}, \& {Ryu}}]{Franchini23}
{Franchini}, A., {Bonetti}, M., {Lupi}, A., {et~al.} 2023, \aap, 675, A100,
  \dodoi{10.1051/0004-6361/202346565}

\bibitem[{{French} {et~al.}(2016){French}, {Arcavi}, \& {Zabludoff}}]{French16}
{French}, K.~D., {Arcavi}, I., \& {Zabludoff}, A. 2016, \apjl, 818, L21,
  \dodoi{10.3847/2041-8205/818/1/L21}

\bibitem[{{French} {et~al.}(2023){French}, {Earl}, {Novack}, {Pardasani},
  {Pillai}, {Tripathi}, \& {Verrico}}]{French23}
{French}, K.~D., {Earl}, N., {Novack}, A.~B., {et~al.} 2023, \apj, 950, 153,
  \dodoi{10.3847/1538-4357/acd249}

\bibitem[{{Gehrels}(1986)}]{Gehrels1986}
{Gehrels}, N. 1986, \apj, 303, 336, \dodoi{10.1086/164079}

\bibitem[{{Giustini} {et~al.}(2020){Giustini}, {Miniutti}, \&
  {Saxton}}]{Giustini20}
{Giustini}, M., {Miniutti}, G., \& {Saxton}, R.~D. 2020, \aap, 636, L2,
  \dodoi{10.1051/0004-6361/202037610}

\bibitem[{{Graur} {et~al.}(2018){Graur}, {French}, {Zahid}, {Guillochon},
  {Mandel}, {Auchettl}, \& {Zabludoff}}]{Graur18}
{Graur}, O., {French}, K.~D., {Zahid}, H.~J., {et~al.} 2018, \apj, 853, 39,
  \dodoi{10.3847/1538-4357/aaa3fd}

\bibitem[{{G{\"u}ltekin} {et~al.}(2009){G{\"u}ltekin}, {Richstone}, {Gebhardt},
  {Lauer}, {Tremaine}, {Aller}, {Bender}, {Dressler}, {Faber}, {Filippenko},
  {Green}, {Ho}, {Kormendy}, {Magorrian}, {Pinkney}, \&
  {Siopis}}]{Gultekin2009}
{G{\"u}ltekin}, K., {Richstone}, D.~O., {Gebhardt}, K., {et~al.} 2009, \apj,
  698, 198, \dodoi{10.1088/0004-637X/698/1/198}

\bibitem[{{Guolo} {et~al.}(2024){Guolo}, {Pasham}, {Zaja{\v{c}}ek}, {Coughlin},
  {Gezari}, {Sukov{\'a}}, {Wevers}, {Witzany}, {Tombesi}, {van Velzen},
  {Alexander}, {Yao}, {Arcodia}, {Karas}, {Miller-Jones}, {Remillard},
  {Gendreau}, \& {Ferrara}}]{Guolo2023}
{Guolo}, M., {Pasham}, D.~R., {Zaja{\v{c}}ek}, M., {et~al.} 2024, Nature
  Astronomy, 8, 347, \dodoi{10.1038/s41550-023-02178-4}

\bibitem[{{G{\"u}ver} \& {{\"O}zel}(2009)}]{Guver09}
{G{\"u}ver}, T., \& {{\"O}zel}, F. 2009, \mnras, 400, 2050,
  \dodoi{10.1111/j.1365-2966.2009.15598.x}

\bibitem[{{Hafen} {et~al.}(2019){Hafen}, {Faucher-Gigu{\`e}re},
  {Angl{\'e}s-Alc{\'a}zar}, {Stern}, {Kere{\v{s}}}, {Hummels}, {Esmerian},
  {Garrison-Kimmel}, {El-Badry}, {Wetzel}, {Chan}, {Hopkins}, \&
  {Murray}}]{Hafen2019}
{Hafen}, Z., {Faucher-Gigu{\`e}re}, C.-A., {Angl{\'e}s-Alc{\'a}zar}, D.,
  {et~al.} 2019, \mnras, 488, 1248, \dodoi{10.1093/mnras/stz1773}

\bibitem[{{Izquierdo-Villalba} {et~al.}(2023){Izquierdo-Villalba}, {Colpi},
  {Volonteri}, {Spinoso}, {Bonoli}, \& {Sesana}}]{Villalba}
{Izquierdo-Villalba}, D., {Colpi}, M., {Volonteri}, M., {et~al.} 2023, \aap,
  677, A123, \dodoi{10.1051/0004-6361/202347008}

\bibitem[{{Johnston} {et~al.}(2008){Johnston}, {Bullock}, {Sharma}, {Font},
  {Robertson}, \& {Leitner}}]{Johnston2008}
{Johnston}, K.~V., {Bullock}, J.~S., {Sharma}, S., {et~al.} 2008, \apj, 689,
  936, \dodoi{10.1086/592228}

\bibitem[{{Kaur} \& {Stone}(2024)}]{Kaur24}
{Kaur}, K., \& {Stone}, N.~C. 2024, arXiv e-prints, arXiv:2405.18500.
\newblock \doarXiv{2405.18500}

\bibitem[{{Keel} {et~al.}(2024){Keel}, {Moiseev}, {Uklein}, \&
  {Smirnova}}]{Keel24}
{Keel}, W.~C., {Moiseev}, A., {Uklein}, R., \& {Smirnova}, A. 2024, \mnras,
  530, 1624, \dodoi{10.1093/mnras/stae946}

\bibitem[{{Keel} {et~al.}(2012){Keel}, {Chojnowski}, {Bennert}, {Schawinski},
  {Lintott}, {Lynn}, {Pancoast}, {Harris}, {Nierenberg}, {Sonnenfeld}, \&
  {Proctor}}]{Keel2012}
{Keel}, W.~C., {Chojnowski}, S.~D., {Bennert}, V.~N., {et~al.} 2012, \mnras,
  420, 878, \dodoi{10.1111/j.1365-2966.2011.20101.x}

\bibitem[{{Keel} {et~al.}(2015){Keel}, {Maksym}, {Bennert}, {Lintott},
  {Chojnowski}, {Moiseev}, {Smirnova}, {Schawinski}, {Urry}, {Evans},
  {Pancoast}, {Scott}, {Showley}, \& {Flatland}}]{Keel2015}
{Keel}, W.~C., {Maksym}, W.~P., {Bennert}, V.~N., {et~al.} 2015, \aj, 149, 155,
  \dodoi{10.1088/0004-6256/149/5/155}

\bibitem[{{Keel} {et~al.}(2017){Keel}, {Lintott}, {Maksym}, {Bennert},
  {Chojnowski}, {Moiseev}, {Smirnova}, {Schawinski}, {Sartori}, {Urry},
  {Pancoast}, {Schirmer}, {Scott}, {Showley}, \& {Flatland}}]{Keel2017}
{Keel}, W.~C., {Lintott}, C.~J., {Maksym}, W.~P., {et~al.} 2017, \apj, 835,
  256, \dodoi{10.3847/1538-4357/835/2/256}

\bibitem[{{Kennedy} {et~al.}(2016){Kennedy}, {Meiron}, {Shukirgaliyev},
  {Panamarev}, {Berczik}, {Just}, \& {Spurzem}}]{Kennedy16}
{Kennedy}, G.~F., {Meiron}, Y., {Shukirgaliyev}, B., {et~al.} 2016, \mnras,
  460, 240, \dodoi{10.1093/mnras/stw908}

\bibitem[{{King}(2022)}]{King22}
{King}, A. 2022, \mnras, 515, 4344, \dodoi{10.1093/mnras/stac1641}

\bibitem[{{Krolik} \& {Linial}(2022)}]{Krolik22}
{Krolik}, J.~H., \& {Linial}, I. 2022, \apj, 941, 24,
  \dodoi{10.3847/1538-4357/ac9eb6}

\bibitem[{{Kroupa}(2001)}]{Kroupa2001}
{Kroupa}, P. 2001, \mnras, 322, 231, \dodoi{10.1046/j.1365-8711.2001.04022.x}

\bibitem[{{Linial} \& {Metzger}(2023)}]{Linial23}
{Linial}, I., \& {Metzger}, B.~D. 2023, \apj, 957, 34,
  \dodoi{10.3847/1538-4357/acf65b}

\bibitem[{{Linial} \& {Metzger}(2024)}]{Linial24}
---. 2024, arXiv e-prints, arXiv:2404.12421, \dodoi{10.48550/arXiv.2404.12421}

\bibitem[{{Linial} \& {Sari}(2023)}]{Linial23a}
{Linial}, I., \& {Sari}, R. 2023, \apj, 945, 86,
  \dodoi{10.3847/1538-4357/acbd3d}

\bibitem[{{Liu} {et~al.}(2023){Liu}, {Malyali}, {Krumpe}, {Homan}, {Goodwin},
  {Grotova}, {Kawka}, {Rau}, {Merloni}, {Anderson}, {Miller-Jones},
  {Markowitz}, {Ciroi}, {Di Mille}, {Schramm}, {Tang}, {Buckley}, {Gromadzki},
  {Jin}, \& {Buchner}}]{Liu23}
{Liu}, Z., {Malyali}, A., {Krumpe}, M., {et~al.} 2023, \aap, 669, A75,
  \dodoi{10.1051/0004-6361/202244805}

\bibitem[{{L{\'o}pez-Cob{\'a}} {et~al.}(2020){L{\'o}pez-Cob{\'a}},
  {S{\'a}nchez}, {Anderson}, {Cruz-Gonz{\'a}lez}, {Galbany}, {Ruiz-Lara},
  {Barrera-Ballesteros}, {Prieto}, \& {Kuncarayakti}}]{LopezCoba2020}
{L{\'o}pez-Cob{\'a}}, C., {S{\'a}nchez}, S.~F., {Anderson}, J.~P., {et~al.}
  2020, \aj, 159, 167, \dodoi{10.3847/1538-3881/ab7848}

\bibitem[{{Lops} {et~al.}(2023){Lops}, {Izquierdo-Villalba}, {Colpi}, {Bonoli},
  {Sesana}, \& {Mangiagli}}]{Lops23}
{Lops}, G., {Izquierdo-Villalba}, D., {Colpi}, M., {et~al.} 2023, \mnras, 519,
  5962, \dodoi{10.1093/mnras/stad058}

\bibitem[{{Lu} \& {Quataert}(2023)}]{Lu22}
{Lu}, W., \& {Quataert}, E. 2023, \mnras, 524, 6247,
  \dodoi{10.1093/mnras/stad2203}

\bibitem[{{Metzger} {et~al.}(2022){Metzger}, {Stone}, \& {Gilbaum}}]{Metzger22}
{Metzger}, B.~D., {Stone}, N.~C., \& {Gilbaum}, S. 2022, \apj, 926, 101,
  \dodoi{10.3847/1538-4357/ac3ee1}

\bibitem[{{Miniutti} {et~al.}(2023){Miniutti}, {Giustini}, {Arcodia}, {Saxton},
  {Read}, {Bianchi}, \& {Alexander}}]{Miniutti23}
{Miniutti}, G., {Giustini}, M., {Arcodia}, R., {et~al.} 2023, \aap, 670, A93,
  \dodoi{10.1051/0004-6361/202244512}

\bibitem[{{Miniutti} {et~al.}(2019){Miniutti}, {Saxton}, {Giustini}, {Alexand
  er}, {Fender}, {Heywood}, {Monageng}, {Coriat}, {Tzioumis}, {Read}, {Knigge},
  {Gandhi}, {Pretorius}, \& {Ag{\'\i}s-Gonz{\'a}lez}}]{Miniutti19}
{Miniutti}, G., {Saxton}, R.~D., {Giustini}, M., {et~al.} 2019, \nat, 573, 381,
  \dodoi{10.1038/s41586-019-1556-x}

\bibitem[{{Pan} {et~al.}(2023){Pan}, {Li}, \& {Cao}}]{Pan23}
{Pan}, X., {Li}, S.-L., \& {Cao}, X. 2023, \apj, 952, 32,
  \dodoi{10.3847/1538-4357/acd180}

\bibitem[{{Patra} {et~al.}(2024){Patra}, {Lu}, {Ma}, {Quataert}, {Miniutti},
  {Chiaberge}, {Filippenko}, \& {Gonz{\'a}lez}}]{Patra24}
{Patra}, K.~C., {Lu}, W., {Ma}, Y., {et~al.} 2024, \mnras,
  \dodoi{10.1093/mnras/stae1146}

\bibitem[{{Pawlik} {et~al.}(2016){Pawlik}, {Wild}, {Walcher}, {Johansson},
  {Villforth}, {Rowlands}, {Mendez-Abreu}, \& {Hewlett}}]{Pawlik16}
{Pawlik}, M.~M., {Wild}, V., {Walcher}, C.~J., {et~al.} 2016, \mnras, 456,
  3032, \dodoi{10.1093/mnras/stv2878}

\bibitem[{{Payne} {et~al.}(2021){Payne}, {Shappee}, {Hinkle}, {Vallely},
  {Kochanek}, {Holoien}, {Auchettl}, {Stanek}, {Thompson}, {Neustadt},
  {Tucker}, {Armstrong}, {Brimacombe}, {Cacella}, {Cornect}, {Denneau},
  {Fausnaugh}, {Flewelling}, {Grupe}, {Heinze}, {Lopez}, {Monard}, {Prieto},
  {Schneider}, {Sheppard}, {Tonry}, \& {Weiland}}]{Payne21}
{Payne}, A.~V., {Shappee}, B.~J., {Hinkle}, J.~T., {et~al.} 2021, \apj, 910,
  125, \dodoi{10.3847/1538-4357/abe38d}

\bibitem[{{Ponti} {et~al.}(2012){Ponti}, {Papadakis}, {Bianchi}, {Guainazzi},
  {Matt}, {Uttley}, \& {Bonilla}}]{Ponti12}
{Ponti}, G., {Papadakis}, I., {Bianchi}, S., {et~al.} 2012, \aap, 542, A83,
  \dodoi{10.1051/0004-6361/201118326}

\bibitem[{{Quintin} {et~al.}(2023){Quintin}, {Webb}, {Guillot}, {Miniutti},
  {Kammoun}, {Giustini}, {Arcodia}, {Soucail}, {Clerc}, {Amato}, \&
  {Markwardt}}]{Quintin23}
{Quintin}, E., {Webb}, N.~A., {Guillot}, S., {et~al.} 2023, \aap, 675, A152,
  \dodoi{10.1051/0004-6361/202346440}

\bibitem[{{Raj} \& {Nixon}(2021)}]{Raj21}
{Raj}, A., \& {Nixon}, C.~J. 2021, \apj, 909, 82,
  \dodoi{10.3847/1538-4357/abdc25}

\bibitem[{{Saxton} {et~al.}(2011){Saxton}, {Read}, {Esquej}, {Miniutti}, \&
  {Alvarez}}]{Saxton11}
{Saxton}, R., {Read}, A., {Esquej}, P., {Miniutti}, G., \& {Alvarez}, E. 2011,
  arXiv e-prints, arXiv:1106.3507.
\newblock \doarXiv{1106.3507}

\bibitem[{{Schawinski} {et~al.}(2015){Schawinski}, {Koss}, {Berney}, \&
  {Sartori}}]{Schawinski15}
{Schawinski}, K., {Koss}, M., {Berney}, S., \& {Sartori}, L.~F. 2015, \mnras,
  451, 2517, \dodoi{10.1093/mnras/stv1136}

\bibitem[{{Schlafly} \& {Finkbeiner}(2011)}]{Schlafly2011}
{Schlafly}, E.~F., \& {Finkbeiner}, D.~P. 2011, \apj, 737, 103,
  \dodoi{10.1088/0004-637X/737/2/103}

\bibitem[{{Schweizer} {et~al.}(2013){Schweizer}, {Seitzer}, {Kelson},
  {Villanueva}, \& {Walth}}]{Schweizer13}
{Schweizer}, F., {Seitzer}, P., {Kelson}, D.~D., {Villanueva}, E.~V., \&
  {Walth}, G.~L. 2013, \apj, 773, 148, \dodoi{10.1088/0004-637X/773/2/148}

\bibitem[{{Sheng} {et~al.}(2021){Sheng}, {Wang}, {Ferland}, {Shu}, {Yang},
  {Jiang}, \& {Chen}}]{Sheng21}
{Sheng}, Z., {Wang}, T., {Ferland}, G., {et~al.} 2021, \apjl, 920, L25,
  \dodoi{10.3847/2041-8213/ac2251}

\bibitem[{{Sniegowska} {et~al.}(2020){Sniegowska}, {Czerny}, {Bon}, \&
  {Bon}}]{Sniegowska20}
{Sniegowska}, M., {Czerny}, B., {Bon}, E., \& {Bon}, N. 2020, \aap, 641, A167,
  \dodoi{10.1051/0004-6361/202038575}

\bibitem[{{Tagawa} \& {Haiman}(2023)}]{Tagawa23}
{Tagawa}, H., \& {Haiman}, Z. 2023, \mnras, 526, 69,
  \dodoi{10.1093/mnras/stad2616}

\bibitem[{{Vazdekis} {et~al.}(2015){Vazdekis}, {Coelho}, {Cassisi},
  {Ricciardelli}, {Falc{\'o}n-Barroso}, {S{\'a}nchez-Bl{\'a}zquez}, {La
  Barbera}, {Beasley}, \& {Pietrinferni}}]{Vazdekis2015}
{Vazdekis}, A., {Coelho}, P., {Cassisi}, S., {et~al.} 2015, \mnras, 449, 1177,
  \dodoi{10.1093/mnras/stv151}

\bibitem[{{Wang} {et~al.}(2024){Wang}, {Ma}, {Wu}, \& {Jiang}}]{Wang2024}
{Wang}, M., {Ma}, Y., {Wu}, Q., \& {Jiang}, N. 2024, \apj, 960, 69,
  \dodoi{10.3847/1538-4357/ad0bfb}

\bibitem[{{Weaver} {et~al.}(2018){Weaver}, {Husemann}, {Kuntschner},
  {Mart{\'\i}n-Navarro}, {Bournaud}, {Duc}, {Emsellem}, {Krajnovi{\'c}},
  {Lyubenova}, \& {McDermid}}]{Weaver18}
{Weaver}, J., {Husemann}, B., {Kuntschner}, H., {et~al.} 2018, \aap, 614, A32,
  \dodoi{10.1051/0004-6361/201732448}

\bibitem[{{Wevers} {et~al.}(2022){Wevers}, {Pasham}, {Jalan}, {Rakshit}, \&
  {Arcodia}}]{Wevers22}
{Wevers}, T., {Pasham}, D.~R., {Jalan}, P., {Rakshit}, S., \& {Arcodia}, R.
  2022, \aap, 659, L2, \dodoi{10.1051/0004-6361/202243143}

\bibitem[{{Wevers} {et~al.}(2023){Wevers}, {Coughlin}, {Pasham}, {Guolo},
  {Sun}, {Wen}, {Jonker}, {Zabludoff}, {Malyali}, {Arcodia}, {Liu}, {Merloni},
  {Rau}, {Grotova}, {Short}, \& {Cao}}]{Wevers23}
{Wevers}, T., {Coughlin}, E.~R., {Pasham}, D.~R., {et~al.} 2023, \apjl, 942,
  L33, \dodoi{10.3847/2041-8213/ac9f36}

\bibitem[{{Wild} {et~al.}(2010){Wild}, {Heckman}, \& {Charlot}}]{Wild2010}
{Wild}, V., {Heckman}, T., \& {Charlot}, S. 2010, \mnras, 405, 933,
  \dodoi{10.1111/j.1365-2966.2010.16536.x}

\bibitem[{{Zabludoff} {et~al.}(1996){Zabludoff}, {Zaritsky}, {Lin}, {Tucker},
  {Hashimoto}, {Shectman}, {Oemler}, \& {Kirshner}}]{Zabludoff1996}
{Zabludoff}, A.~I., {Zaritsky}, D., {Lin}, H., {et~al.} 1996, \apj, 466, 104,
  \dodoi{10.1086/177495}

\bibitem[{{Zhao} {et~al.}(2022){Zhao}, {Wang}, {Zou}, {Wang}, \&
  {Dai}}]{Zhao22}
{Zhao}, Z.~Y., {Wang}, Y.~Y., {Zou}, Y.~C., {Wang}, F.~Y., \& {Dai}, Z.~G.
  2022, \aap, 661, A55, \dodoi{10.1051/0004-6361/202142519}

\bibitem[{{Zhou} {et~al.}(2024){Zhou}, {Zhong}, {Zeng}, {Huang}, \&
  {Pan}}]{Zhou24}
{Zhou}, C., {Zhong}, B., {Zeng}, Y., {Huang}, L., \& {Pan}, Z. 2024, arXiv
  e-prints, arXiv:2405.06429, \dodoi{10.48550/arXiv.2405.06429}

\end{thebibliography}
\bibliographystyle{aasjournal}
\end{document}